\documentclass[conference]{IEEEtran}
\IEEEoverridecommandlockouts

\usepackage{cite}
\usepackage{amsmath,amssymb,amsfonts}
\usepackage{algorithmic}
\usepackage{graphicx}
\usepackage{hyperref}
\usepackage{textcomp}
\usepackage[table]{xcolor}
\usepackage{booktabs}

\usepackage{subcaption}
\usepackage{cleveref}
\usepackage{paralist}
\usepackage{enumitem}
\usepackage{multirow}
\usepackage[multiple]{footmisc}

\usepackage{pifont}%

\usepackage[show]{chato-notes} %

\usepackage[]{mdframed}

\usepackage{array}
\newcolumntype{C}[1]{>{\centering\let\newline\\\arraybackslash\hspace{0pt}}m{#1}}

\title{SoK: Pitfalls in Evaluating Black-Box Attacks}

\makeatletter
\newcommand{\linebreakand}{%
  \end{@IEEEauthorhalign}
  \hfill\mbox{}\par
  \mbox{}\hfill\begin{@IEEEauthorhalign}
}
\makeatother
\author{
\IEEEauthorblockN{Fnu Suya\IEEEauthorrefmark{1}}
\IEEEcompsocitemizethanks{\IEEEcompsocthanksitem\IEEEauthorrefmark{1}Both authors contributed equally.}
\IEEEauthorblockA{\textit{University of Maryland College Park} \\
suya@umd.edu}
\and
\IEEEauthorblockN{Anshuman Suri\IEEEauthorrefmark{1}}
\IEEEauthorblockA{\textit{University of Virginia} \\
anshuman@virginia.edu}
\and
\IEEEauthorblockN{Tingwei Zhang}
\IEEEauthorblockA{\textit{Cornell University} \\
tz362@cornell.edu}
\and
\IEEEauthorblockN{Jingtao Hong}
\IEEEauthorblockA{\textit{Columbia University} \\
jh4760@columbia.edu}
\linebreakand %
\IEEEauthorblockN{Yuan Tian}
\IEEEauthorblockA{\textit{University of California Los Angeles} \\
yuant@ucla.edu}
\and
\IEEEauthorblockN{David Evans}
\IEEEauthorblockA{\textit{University of Virginia} \\
evans@virginia.edu}
}

\newcommand{\revision}[1]{#1}

\begin{document}
\maketitle

\newcommand{\codelink}{\url{https://github.com/iamgroot42/blackboxsok}}

\newcommand{\bm}[1]{\boldsymbol{#1}}

\newcommand{\yuan}[1]{\textcolor{orange}{\bf \emph{Yuan: #1}}}
\newcommand{\snote}[1]{\mynote[author=Suya]{#1}}
\newcommand{\anote}[1]{\mynote[author=Anshuman]{#1}}
\newcommand{\dnote}[1]{\textcolor{blue}{\bf \emph{Dave: #1}}}
\newcommand{\anoteinline}[1]{\textcolor{magenta}{\bf \emph{Anshuman: #1}}}
\newcommand{\tnote}[1]{\textcolor{cyan}{\bf \emph{Tingwei: #1}}}
\newcommand{\scott}[1]{\textcolor{violet}{\bf \emph{Scott: #1}}}
\newcommand{\imark}{\textbf{$\dagger$}}
\newcommand{\mware}[1]{\textcolor{orange}{#1}}
\newcommand{\shortgap}{\vspace{6pt}}
\newcommand\shortsection[1]{\vspace{6pt}{\noindent\bf #1.}}
\newcommand\shortersection[1]{\vspace{6pt}{\noindent\em #1.}}
\newcommand{\cmark}{\ding{51}}
\newcommand{\xmark}{\ding{55}}

\newcommand{\fgsm}{FGSM}
\newcommand{\ifgsm}{I-FGSM}
\newcommand{\admixfgsm}{Admix-FGSM}
\newcommand{\mifgsm}{MI-FGSM}
\newcommand{\nifgsm}{NI-FGSM}
\newcommand{\vnifgsm}{VNI-FGSM}
\newcommand{\vmifgsm}{VMI-FGSM}
\newcommand{\emifgsm}{EMI-FGSM}
\newcommand{\midifgsm}{MIDI-FGSM}
\newcommand{\smimifgsm}{SMIMI-FGSM}
\newcommand{\smifgsm}{SMI-FGSM}
\newcommand{\squarenew}{Hybrid-Square}
\newcommand{\topksquare}{Square: top-\emph{k}}
\newcommand{\odsrgf}{ODS-RGF}

\newcommand{\niterstargeted}{{40}\xspace}

\newcommand{\emptysymbol}{$\varnothing$}
\newcommand{\emptysymbolcell}{\cellcolor{red!20}\emptysymbol}

\newcommand{\topk}{\mbox{top-\emph{k}}}

\newcommand{\figscale}{1.0}

\newcommand{\ie}{{i.e.,}\xspace}
\newcommand{\eg}{{e.g.,}\xspace}
\newcommand{\cf}{{cf.}\xspace}
\newcommand{\etal}{{et al.}\xspace}
\newcommand{\vs}{{vs.}\xspace}
\newcommand{\etc}{{etc.}\xspace}

\begin{abstract}
Numerous works study black-box attacks on image classifiers, where adversaries generate adversarial examples against unknown target models without having access to their internal information.
However, these works make different assumptions about the adversary's knowledge, and current literature lacks cohesive organization centered around the threat model. To systematize knowledge in this area, we propose a taxonomy over the threat space spanning the axes of feedback granularity, the access of interactive queries, and the quality and quantity of the auxiliary data available to the attacker. Our new taxonomy provides three key insights.
1) Despite extensive literature, numerous under-explored threat spaces exist, which cannot be trivially solved by adapting techniques from well-explored settings. We demonstrate this by establishing a new state-of-the-art in the less-studied setting of access to \topk\ confidence scores by adapting techniques from well-explored settings of accessing the complete confidence vector but show how it still falls short of the more restrictive setting that only obtains the prediction label, highlighting the need for more research.
2) Identifying the threat models for different attacks uncovers stronger baselines that challenge prior state-of-the-art claims. We demonstrate this by enhancing an initially weaker baseline (under interactive query access) via surrogate models, effectively overturning claims in the respective paper.
3) Our taxonomy reveals interactions between attacker knowledge that connect well to related areas, such as model inversion and extraction attacks. We discuss how advances in other areas can enable stronger black-box attacks. 
Finally, we emphasize the need for a more realistic assessment of attack success by factoring in local attack runtime. This approach reveals the potential for certain attacks to achieve notably higher success rates. We also highlight the need to evaluate attacks in diverse and harder settings and underscore the need for better selection criteria when picking the best candidate adversarial examples.
\end{abstract}

\section{Introduction}

Machine learning models, including models using deep learning, are well known to be vulnerable to specially-crafted inputs, known as \emph{adversarial examples} (AEs), that are designed to induce incorrect predictions. Most early studies of adversarial examples focused on white-box settings where the adversary has full access to the target model~\cite{szegedy2013intriguing,goodfellow2014explaining}. Black-box settings consider scenarios where the adversary has limited access to the target model. Such settings are a more practical threat to many deployed systems~\cite{ilyas2018black,bhagoji2018practical,apruzzese2023real} where the model is not revealed directly. In these attacks, known as \emph{black-box} or \emph{API-only} attacks, the adversary can interact with the target model using API queries but does not have direct access to the model's parameters and may have varying degrees of knowledge about the model architecture, training data, and training process.
Previous surveys of such attacks~\cite{bhambri2019survey,mahmood2021back}
categorize representative attacks based on their adopted methods but overlook differences in assumptions about the adversary's knowledge and capabilities. These assumptions can vary wildly, depending on the resources available for the attacker and the kind of access to the model the API provides. Different assumpations have a significant impact on what attacks are possible in practice. Furthermore, attack evaluations typically rely solely on attack success rates (and query cost for interactive attacks), ignoring how attack success varies across different examples and tasks. This disconnect makes it hard to map out the threat space, leading to improper evaluation of attacks and limiting our understanding of the actual threats.

\shortsection{Contributions}
\revision{We started by surveying black-box attacks on image classifiers published in major security (Usenix Security, IEEE S\&P, CCS, NDSS), machine-learning (ICML, NeurIPS, ICLR, KDD, AAAI, IJCAI) and computer vision (CVPR, ICCV, ECCV) venues. In particular, we identified relevant papers published in the aforementioned top-tier conferences by searching with keywords ``transfer'', ``attack'', ``black-box'', and ``query'' from the year 2014 (the year of the first paper \cite{szegedy2013intriguing} on generating adversarial examples on deep neural networks) to 2023. In addition to these works, we conducted a thorough search of papers referenced within them and of relevant works citing these identified papers, covering both peer-reviewed papers and preprints online with the best effort.} This leaves us with 164 attacks, of which 102 are published in major security and machine learning conferences. With the surveyed attacks, we propose a new taxonomy for existing black-box attacks, organized around assumptions on their threat models. Our taxonomy spans  four dimensions (\autoref{sec:tax_dims}): 
\begin{inparaenum}[1)]
    \item  interactive queries to the target model allowed,
    \item information provided by the target model's API,
    \item quality of the initial auxiliary data available to the adversary, and         
    \item quantity of the initial auxiliary data available to the adversary.
\end{inparaenum} 
\revision{These dimensions are chosen based on the underlying components that enable successful black-box attacks in practice---the feedback available for the attackers to adjust the strategy (whether interactive queries are permitted, and the granularity of the feedback provided if any) and the resources attackers can leverage (quantity and quality of data initially available for the attacker, as well as the availability of pretrained models online, independent from the auxiliary data)} We categorize the existing literature using our proposed taxonomy (\autoref{sec:attack category threat}), focusing on image classifiers as the most widely studied domain. Our observations result in three key findings:
\begin{enumerate} %
    \item Most prior works are concentrated in specific regions of the taxonomy, with several important and practically relevant settings that have not been well explored. Much of this knowledge gap is also likely a technical gap, and we demonstrate this with preliminary experiments on devising stronger baselines in one of the under-explored settings. Despite establishing state-of-the-art attack success, many methods fall short of attacks from more restrictive but well-explored settings, reinforcing the importance of investing research into these under-explored areas (\autoref{sec:unexplored areas}).
    \item Some works propose new attacks and compare them to existing baseline attacks under threat models \revision{more restrictive than their own}, which can underestimate the potency of baselines given enough knowledge. We empirically demonstrate how claims of methods outperforming previous ones can often be invalidated when prior attacks are adapted to and compared under the same threat model (\autoref{sec:stronger_baselines}).
    \item A closer look at the threat space reveals the scope for utilizing available resources in different and potentially better ways. In particular, attackers with access to some initial auxiliary data and pre-trained models may leverage advances in model extraction~\cite{tramer2016stealing,jagielski2020high} and model inversion attacks~\cite{fredrikson2015model, wang2021improving} to enable stronger attacks. We discuss the possible usage of this interaction and motivate future research along this direction (\autoref{sec:new direction}).
\end{enumerate}
Transfer attack evaluations in the literature focus on the number of local optimization iterations as a normalizing factor when comparing attacks. While well intended, such measures are misaligned with practical adversaries' goals: picking an attack that maximizes success within some given time frame. Our evaluation of transfer attacks 1) shows how normalizing for time allows some attacks to run for more iterations and achieve higher success rates; 2) motivates future research to work on better metrics to select better local candidates of adversarial examples, and to evaluate attacks in diverse and harder attack settings.
We clarify that adversaries may conduct training~\cite{mehra2021robust,radiya2021data} with prediction-time attacks. While such adversaries can be extremely potent, our current taxonomy focuses on prediction-time attacks and thus does not capture dynamically changing, possibly poisoned, target models~\cite{diochnos2020lower}.

To support comprehensive evaluations of attacks and defenses, we provide a modular codebase at \codelink\ 

\section{Background} \label{sec:background}
We first introduce background on adversarial examples (\autoref{sec:ae background}), and then review related works (\autoref{sec:related work}).
\subsection{Introduction of Adversarial Examples}\label{sec:ae background}
In image classification tasks, given a model/classifier $f$ that takes input $\bm{x}$ (with ground truth label $c(\bm{x})$) and generates a prediction $f(\bm{x})$, the goal of adversary is to achieve some attack goals by adding an (imperceptible) bounded perturbation $\bm{\delta}$ onto $\bm{x}$. Depending on the attack goals, there can be \emph{untargeted} and \emph{targeted} attack goals. Untargeted attacks aim to induce a predicted class on the perturbed input $\bm{x}+\bm{\delta}$ that is different from $c(\bm{x})$, namely, $f(\bm{x}+\bm{\delta}) \neq c(\bm{x}).$ Note that we define the attack goal of misclassification with respect to the ground-truth label $c(\bm{x})$ of input $\bm{x}$, which is consistent with implicit assumptions made in the surveyed black-box attacks (i.e., the evaluations consider misclassifying correctly labeled samples and assume $c(\bm{x})=c(\bm{x}+\bm{\delta})$). However, there can be other definitions of untargeted attacks that are more related to the definition of the adversarial risk of a model $f$, such as causing misclassification with respect to $f(\bm{x})$ or $c(\bm{x}+\bm{\delta})$ (if different from $c(\bm{x})$). Diochnos et al.~\cite{diochnos2018adversarial} provide a more detailed comparison between these definitions, but they are the same for the setting considered in this paper. Targeted attacks ensure the perturbed sample $\bm{x}+\bm{\delta}$ is misclassified into a particular label $\hat{y}$ that is in the interest of the adversary, namely, $f(\bm{x} + \bm{\delta}) = \hat{y}.$
The bounded perturbation is $\bm{\delta}$ is constrained by some perturbation budget $\epsilon$ to avoid raising suspicion, although some works also consider minimizing the perturbation magnitude \cite{chen2017zoo,carlini2017towards}. The most common constraint is to limit the $\ell_p$ norm of the perturbation $\bm{\delta}$, namely $\|\bm{\delta}\|_p\leq \epsilon$. 

The white-box attacks have access to all the internal information of the target model and therefore, can optimize the perturbation $\bm{\delta}$ with respect to (w.r.t) some loss function (e.g. maximize cross-entropy w.r.t $c(\bm{x})$ in untargeted and maximize the loss w.r.t $\hat{y}$ in targeted settings) to generate the adversarial examples using gradient descent \cite{goodfellow2014explaining}. In contrast, black-box attacks do not have access to the model's internal information and, therefore, either rely on transfer attacks if some local surrogates are available (\autoref{sec:transfer attacks}) or black-box optimization if interactive access is permitted (\autoref{sec:query access}).

\subsection{Related Works}
\label{sec:related work}
\shortsection{Surveys on Black-box Attacks} Two survey papers already cover black-box attacks in the vision domain~\cite{bhambri2019survey,mahmood2021back}. These papers categorize representative attacks by methods, identifying the best attacks and offering meta-analyses of their reported results. However, they draw conclusions from experimental results reported in prior works, which are spread across incompatible settings and threat models. 
In contrast to these works, we provide the taxonomy based on the threat model, which enables a better understanding of how attacks relate and how they should be compared. This, in turn, allows us to evaluate attacks in consistent test environments and draw meaningful conclusions.
The most relevant previous work is Zhao et al.'s comprehensive evaluation of transfer attacks in the image domain~\cite{zhao2022towards}. They focus on understanding the robustness of different defenses against untargeted transfer attacks at a fixed perturbation budget and compare the visual stealthiness of different attacks with the same norm constraint. In contrast, we focus on general black-box attacks and compare attacks across various threat models.

\shortsection{Relevant SoKs} There are several previous SoK papers on adversarial machine learning, focusing on different topics ranging from categorizing attacks on audio recognition systems~\cite{abdullah2021sok} to certified robustness for adversarial examples~\cite{li2020sok}. Papernot et al.\ provide a general systematization of adversarial machine learning, but \revision{do not focus} on black-box attacks~\cite{papernot2018sok}. Carlini et al.~\cite{carlini2019evaluating} provide a set of guidelines for proper evaluation of adversarial robustness \revision{in white-box settings}. While some recommendations, such as proper threat model categorization and running attacks until convergence, apply to black-box attacks as well, \revision{we provide a concrete taxonomy with detailed analyses and advocate for time-based comparison of attacks.} Our paper is the first to systematize knowledge of black-box attacks based on their applicable threat models. 
\section{Attack Methods}
\label{sec:method category}

Most attacks use the same underlying principles and
structure, but make advances in one or more aspects of the attack process. In this section, we categorize attacks based on their strategies, building on top of the categories provided in the prior literature \cite{zhao2022towards,mahmood2021back}. These categories help better understand similarities and connections between existing attacks and identify scope for improvement and combinations of advancements.
This categorization is orthogonal to the threat-model based taxonomy we introduce in \autoref{sec:tax_dims}.

\subsection{Transfer attacks}
\label{sec:transfer attacks}
Transfer attacks first generate adversarial examples for local surrogate models with white-box access and then attempt to transfer those local adversarial examples to the target model \cite{goodfellow2014explaining,kurakin2016adversarial}. The success of a transfer attack depends on how similar (at least with respect to the relevant decision boundaries) the local models are to the target model and how effective the local attack is at finding generalizable adversarial examples against the local models. We mainly adopt the terminology used by Zhao et al.~\cite{zhao2022towards} to describe existing attacks. However, we added a category of \emph{Better Loss Functions}, which customizes the loss function for better transferability.  

\shortsection{Gradient Stabilization}
The idea behind the gradient stabilization is to make the model less prone to overfitting to the local model and improve the transferability to the unknown target model, through utilizing the spatial~\cite{wang2022enhancing,wu2018understanding,gao2020patch,gao2020patch+,li2020regional,gao2021staircase,tan2022improving,wang2021feature,fang2022learning} 
and temporal correlation~\cite{dong2018boosting,lin2019nesterov,zou2022making,wang2021boosting,Jang_2022_CVPR,wang2021enhancing,lu2021towards,wang2021feature,he2022boosting,tan2022improving} among the gradients.%

\shortsection{Input Augmentation} The main idea is to augment and diversify inputs to increase the transferability of adversarial examples to different target models, and is similar to generating more generalizable models through augmenting the training data \cite{zhao2022towards,liang2021uncovering}. Early works applied common and hand-crafted input augmentations~\cite{xie2019improving,zou2020improving,yang2022adversarial,dong2019evading,lin2019nesterov,Jang_2022_CVPR,long2022frequency,fang2022learning,Byun_2022_CVPR,wang2023structure}. Recent works have focused on learning better input transformations with neural networks \cite{wu2021improving} or finding the best input from existing ones using neural networks \cite{yuan2021adaptive,zhang2023improving} or reinforcement learning~\cite{yuan2021automa}.

\shortsection{Better Loss Functions}
Recently, more research is focused on leveraging intermediate model features~\cite{li2020towards,zhao2021success,qin2021boosting,wang2020unified,ganeshan2019fda,inkawhich2020transferable,inkawhich2020perturbing,huang2019enhancing,li2020yet,liu2019s,zhou2018transferable,inkawhich2019feature,wang2021feature,wu2020boosting,zhang2022improving,wang2021exploring} instead of the model outputs, with an intuition that intermediate feature representations may be more generic and transferable~\cite{kornblith2019similarity,yosinski2014transferable}, which can be further enhanced with interpretability techniques to focus on relevant features~\cite{selvaraju2017grad,sundararajan2017axiomatic}.

\shortsection{Surrogate Refinement}
Different local surrogate models can have different transferability to the target model, and the most naiv\"e approach to improve transferability is to adopt an ensemble of models\cite{liu2016delving,li2020learning}. Other techniques focus on improving transferability of a single surrogate, such as modifying for or using better local architectures (\eg using skip connections) \cite{wu2020skip,duan2022adversarial,Xu_2023_ICCV}; modifying the activations functions \cite{zhu2021rethinking,zhang2021backpropagating,guo2020backpropagating,wang2023rethinking}; modifying the training strategy (\eg adopting adversarial training, early stopping) \cite{salman2020adversarially,utrera2020adversarially,deng2021adversarial,zhou2022improving,springer2021little,zhang2021early,yang2022boosting}; identifying proper source models with meta-learning \cite{yuan2021meta,fang2022learning}. 

\shortsection{Generative Models}
Unlike iterative attacks, existing works also train generative models that, given an input, produce the corresponding adversarial example. Existing works focus on using better loss functions to train the generators \cite{poursaeed2018generative,naseer2019cross,kanth2021learning,zhang2022beyond} and also using better generator architectures such as class-specific \cite{naseer2021generating} or class conditional ones \cite{yang2022boosting}.

\subsection{Query-based attacks}
\label{sec:query attacks}
Query-based attacks refine the candidate adversarial examples with interactive queries until the attacker's objective is achieved. Below, we introduce the common methodologies \revision{which are based on gradient-estimation or gradient-free attacks. Gradient-free attacks can apply to much broader settings, especially for non-differentiable models, whereas gradient-estimation methods are illogical for such models. While estimating a non-differential target model with certain differentiable approximations is possible, this remains an open question. For deep neural networks, gradient estimation methods perform better on tasks that involve minimizing the perturbation magnitude \cite{zhang2021progressive,li2021nonlinear}, while gradient-free attacks tend to work better under fixed perturbation budgets, especially for the $\ell_{\infty}$-norm \cite{andriushchenko2020square,chen2020rays}.}

\subsubsection{Gradient-estimation Attacks}
These attacks work by estimating the gradients of the unknown target model and updating candidate adversarial examples accordingly. This technique can be applied irrespective of the target models returning full prediction scores \cite{chen2017zoo,ilyas2018black} or just the prediction class \cite{cheng2018query,cheng2019sign,chen2020hopskipjumpattack}. 

\shortsection{Complete Confidence Vector}
In this setting, confidence scores of all classes are available, and attacks start from the original seed image and gradually search for better perturbations with the estimated gradients. Ever since the first work on estimating the gradient for every coordinate with finite-difference method \cite{chen2017zoo}, subsequent works focused on finding more efficient gradient estimation strategies, mostly by finding a better random perturbation vector to estimate the gradient efficiently for the finite-difference method ~\cite{du2018towards,ilyas2018black,liu2018signsgd,uesato2018adversarial,bhagoji2018practical,li2020projection,ilyas2018prior,shi2019curls,zhao2020towards,zhao2019design,chen2020frank,al2019sign,li2019nattack}. 

\shortsection{Hard-Label}
The hard-label attacks are more restrictive and can only access the prediction label of the highest confident class; hence, the attacks usually require a reference image that satisfies the attacker's objective (\eg the reference image is from the intended target class for misclassification) to generate a likely-to-succeed perturbation and then focus on minimizing the size of the perturbation (measured by $\ell_p$-norm such as $\ell_2$) with the estimated gradients. Since the first work \cite{cheng2018query}, various techniques are proposed to improve the gradient estimation quality and boost attack performance \cite{cheng2019sign,chen2020hopskipjumpattack,simon2021popskipjump,wang2022query,rahmati2020geoda,liu2019geometry,zhao2021improved}. 

\subsubsection{Gradient-free Attacks}
s the name suggests, gradient-free attacks do not rely on estimating the target model gradients. These attacks are diverse in terms of their methodologies.

\shortsection{Complete Confidence Vector}
Gradient-free attacks with complete confidence vector range from classical black-box optimization techniques (\eg genetic algorithms, evolution strategies, Bayesian Optimization)~\cite{chen2019poba,alzantot2019genattack,meunier2019yet,ru2019bayesopt,suya2017query} to efficient random search strategies~\cite{moon2019parsimonious,narodytska2016simple,guo2019simple,andriushchenko2020square,tran2022exploiting,croce2019sparse,croce2022sparse,shiva2017simple}. The key is to find an effective low-dimensional subspace to generate perturbations and then map back to the original input space. The recent efficient random search-based attacks \cite{andriushchenko2020square,guo2019simple} are the current state-of-the-art to generate norm-bounded perturbations.

\shortsection{Hard-Label}
The first type of gradient-free methods are based on random walk with various sampling distributions~\cite{brendel2017decision,brunner2019guessing,dong2019efficient,shi2020polishing,li2021aha,sun2022query} or directions based on the geometry of the decision boundary~\cite{maho2021surfree}. Recently, more efficient attacks are proposed using diverse techniques such as random search~\cite{chen2020rays,chen2020boosting,midtlid2022magnitude}, evolution strategies~\cite{vo2022query} or utilization of geometric properties of the boundary~\cite{wang2022triangle}. For norm-constrained adversaries, especially in $\ell_\infty$-norm, the random search-based methods achieve the state-of-the-art performance \cite{chen2020rays,chen2020boosting}.

\subsection{Hybrid Attacks}\label{sec:hybrid attacks}
These attacks utilize surrogate models, like transfer attacks, and submit queries to the target model. We name these attacks ``hybrid attacks" to distinguish them from pure transfer or query-based attacks. There are mainly two types of hybrid attacks. The first type leverages surrogate models to enhance query-based attacks by providing better starting points (i.e., warm starting)~\cite{suya2020hybrid} or providing better sampling space of perturbation~\cite{huang2019black,tashiro2020diversity,yang2020learning,lord2022attacking,cheng2019improving,guo2019subspace,ma2020switching} for the query-based attacks. The second type improves available \emph{surrogate models} with labeled queries from the target model, including fine-tuning the models~\cite{suya2020hybrid,yang2020learning,chen2022optimized,chen2021querynet} or finding proper weights for individual models in the model ensemble~\cite{cai2022blackbox}, so that the transferability from these similar models can be significantly improved in the later stage. The only exception from above is that queries from the target model can also be combined with local explanation techniques~\cite{ribeiro2016should} to select the most transferable single model from a set of classifiers~\cite{severi2022bad}.
 
\iffalse
\todo{Introduce current white-box attacks in a short paragraph to give readers overview of how these attacks work}

\fgsm~\cite{goodfellow2014explaining}, \ifgsm~\cite{kurakin2018adversarial}, \mifgsm~\cite{dong2018boosting}, \nifgsm~\cite{lin2019nesterov}, VMI-FGSM~\cite{wang2021enhancing}, VNI-FGSM~\cite{wang2021enhancing}, \smifgsm~\cite{wang2022enhancing}, \emifgsm~\cite{wang2021boosting}, I-FGSSM~\cite{gao2021staircase}, DI-FGSM~\cite{xie2019improving}, TI-FGSM~\cite{dong2019evading}, SI-FGSM~\cite{lin2019nesterov}, \admixfgsm~\cite{wang2021admix}, ODS-FGSM~\cite{tashiro2020diversity}
\fi

\section{Taxonomy Threat Model}
\label{sec:tax_dims}

We propose a new attack taxonomy organized around the threat model assumptions of an attack, using four separate dimensions to categorize assumptions made by each attack. Within each dimension, we describe different categories in order of increasing knowledge available to the adversary (\autoref{sec:query access} - \autoref{sec:quant aux data}). We then discuss the existence of pretrained models as a sub-axis \autoref{sec:pretrained models} and how it may interact with the main axes of our dimension. 
We then use our taxonomy to categorize attacks (\autoref{sec:attack category threat}) and report our insights with directions for future research (\autoref{sec:taxonomy_insights}).

\subsection{Query Access}\label{sec:query access}
Query access captures the adversary's ability to query the target model \textit{before} sending its final adversarial input. We group access levels into two characteristic settings: 
\begin{enumerate}[label=(\alph*)]
    \item \textbf{No Interactive Access:} 
    the adversary has absolutely no opportunity to query the target model interactively. Likely scenarios include situations where the adversary has only one-way communication with the target model through an indirect victim. For example, the adversary may want to generate malware that bypasses the victim's malware classification system but without any way to query that system directly. This is the most challenging attack setting where the adversary has no opportunity to learn from feedback from the target model.
    
    \item \textbf{With Interactive Access:} 
    a more relaxed setting and still has wide applications in practice. In this setting, the attacker can interactively query the target model and adjust subsequent queries by leveraging its history of queries. However, the number of queries that can be submitted might be constrained significantly in practical cases,\eg rate limits imposed by the target model API, the financial cost involved in making queries, or simply the attackers wanting to avoid raising suspicion. In other situations, the attackers may still be able to query the target model as often as they wish. The most concrete example of unlimited black-box query access would be one where the adversary has access to the model on their hardware, but it is encrypted in a secure enclave (\eg Intel SGX as the Trusted Execution Environment) that protects its parameters\cite{hou2021model,gu2018yerbabuena}.

\end{enumerate}

\subsection{API Feedback}
This dimension captures the granularity of information the target model's API returns for a given query. We break this down into three distinct categories:
\begin{enumerate}[label=(\alph*)]
    \item \textbf{Hard-Label:} the only value returned by the API is the predicted label for the given query input. For instance, a face-recognition based utility may only provide a label for match/mismatch.
    \item \textbf{Top-K:} the model API returns confidence scores for the top-k ($1\leq k < N$, for $N$ classes) labels. This aligns well with most real-world predictive APIs, which often return confidence values for a few most likely classes to minimize network overhead. This setting provides more information than hard-label access even when $k=1$, since the confidence score for the predicted label is made available. For example, Google's Cloud Vision API\footnote{\url{https://cloud.google.com/vision/docs/labels}} uses labels from their Knowledge Graph API\footnote{\url{https://developers.google.com/knowledge-graph/reference/rest/v1/}}, which has tens of thousands of labels, and returning classification scores for all classes is unlikely to be helpful for benign users.
    \item \textbf{Complete Confidence Vector:} the API returns confidence scores for all classes. This may correspond to the enclave-based setting described above, or one where the number of classes is low enough for an API to return all related information. %
\end{enumerate}
Below, we describe auxiliary information available to attackers for more efficient attacks. We define two axes of 1) the \emph{quality} of data and 2) the \emph{quantity} of data. 

\subsection{Quality of Initial Auxiliary Data} \label{sec:qual aux data}
This dimension captures the correctness of the adversary's priors on the target model's training data. Higher quality of auxiliary data indicates that the attackers can conduct the attack without considering potential distributional gaps. \revision{In this paper, we capture such distributional gaps using the overlap between the feature or label space of two distributions (corresponding to the target model's data and auxiliary data). Feature space overlap refers to same/similar samples in the data feature (\eg images of dogs in two distributions) regardless of the assigned labels (\eg different labels for the same image, depending on different tasks). We discuss overlap on distributions, not on datasets, because distributions are more fundamental than the (sampled) datasets.}

\begin{enumerate}[label=(\alph*)]
\item \textbf{No Overlap:} auxiliary data available to the adversary does not overlap in the data features and the labels. This setting is closest to real-world APIs, where knowledge about the target model's training data is obfuscated and often proprietary (like GPT-4). %

\item \textbf{Partial Overlap:} auxiliary data available to the adversary has partial overlaps (in the distributional sense) with the private training data of the target model regarding data features or labels. This setting best matches scenarios where the training data of the target model includes some publicly available datasets.

\item \textbf{Complete Overlap} auxiliary data available to the attacker is the same as the target model's training data, or sampled from the same underlying distribution (\revision{\ie same label space and feature space}). For example, the target model could be trained on a publicly available dataset, and this information may be public.

\end{enumerate}
\revision{Notably, removing the high overlap in data distributions can significantly undermine the attack success~\cite{richards2021adversarial}. The authors propose a variant of PGD (masked PGD) to mitigate the performance degradation due to distributional gap.}

\renewcommand\arraystretch{1.1}
\begin{table*}
\centering
\scalebox{0.85} 
{
\begin{tabular}
{C{0.5cm}C{1.5cm}C{5.5cm}C{4.5cm}C{0.75cm}C{5.5cm}}%
\toprule
\multirow{2}{*}{Quality} & \multirow{2}{*}{Quantity} & \multirow{2}{*}{No Interactive Access} &  \multicolumn{3}{c}{With Interactive Access} \\
 & &
& Hard-Label & Top-K & Complete Confidence Vector \\ \midrule
\multirow{5}{*}{\rotatebox[origin=c]{90}{\quad\quad\quad None}}
& Insufficient & 
Frequency Manipulation~\cite{zhang2022practical}\newline
\textbf{w/ Pretrained Surrogate$^*$:}\newline
Better Loss:~\cite{huan2020data,inkawhich2021can,kanth2021learning,lu2020enhancing,mopuri2018generalizable,mopuri2017fast,naseer2019cross,naseer2018task,qin2021adversarial,wu2020decision,zhang2021data,zhang2022beyond,richards2021adversarial}\newline
Better Loss for AE Generator: \cite{kanth2021learning,naseer2019cross,naseer2018task}
 & 
Random walk:\cite{brendel2017decision,brunner2019guessing,dong2019efficient,li2021aha,shi2020polishing,sun2022query,maho2021surfree} \newline
Gradient estimation: %
\cite{cheng2018query,cheng2019sign,chen2020hopskipjumpattack,liu2019geometry,simon2021popskipjump,wang2022query,rahmati2020geoda,zhao2021improved}\newline
Other Gradient-free:
\cite{chen2020rays,chen2020boosting,midtlid2022magnitude,vo2022query,wang2022triangle}\newline
Classic Black-box Opt.:\cite{shukla2021simple,zhao2019design}\newline
 & NES~\cite{ilyas2018black} & 
{Gradient Estimation:} %
\cite{chen2017zoo,al2019sign,chen2020frank,bhagoji2018practical,du2018towards, ilyas2018black,ilyas2018prior,li2020projection,li2019nattack, liu2018signsgd,shi2019curls,uesato2018adversarial,zhao2020towards, zhao2019design}\newline
{Classic Black-box Opt.:} %
\cite{chen2019poba,alzantot2019genattack,meunier2019yet,ru2019bayesopt,suya2017query}\newline
{Efficient Random Search:}\newline
\cite{chen2019poba,alzantot2019genattack,croce2022sparse,croce2019sparse,guo2019simple,moon2019parsimonious,meunier2019yet,narodytska2016simple,tran2022exploiting,shiva2017simple,andriushchenko2020square}
  \\ \cmidrule{2-6}

\multicolumn{1}{c}{} & Sufficient & \emptysymbolcell & \emptysymbolcell & \emptysymbolcell & \emptysymbolcell \\
\midrule
\multirow{4}{*}{\rotatebox[origin=c]{90}{\quad\quad Partial}} & Insufficient & 
\textbf{w/ Pretrained Surrogate$^*$:}\newline
Better Loss: \cite{inkawhich2021can,qin2021adversarial,zhang2022beyond,richards2021adversarial}
 & \emptysymbolcell & \emptysymbolcell & Boost Existing Methods w/ Trained Generator: \cite{yatsura2021meta} \\ \cmidrule{2-6}
 & Sufficient & \emptysymbolcell & \emptysymbolcell & \emptysymbolcell & \emptysymbolcell \\ \midrule
\multirow{11}{*}{\rotatebox[origin=c]{90}{\quad\quad\quad\quad\quad\quad Complete}} & Insufficient & 
Train Shallow Surrogate: \cite{li2020practical,sun2022towards}\newline
\textbf{w/ Pretrained Surrogate$^*$:}\newline
(Basic) Gradient Sign: \cite{goodfellow2014explaining,kurakin2016adversarial}\newline
Input Augmentation:
\cite{dong2019evading,Jang_2022_CVPR,lin2019nesterov,long2022frequency,wang2021admix,wu2021improving,xie2019improving,yang2022adversarial,yuan2021adaptive,yuan2021automa,zou2020improving,Byun_2022_CVPR,wang2023structure,zhang2023improving,fang2022learning}
\newline

Gradient Stabilization:
\cite{wang2022enhancing,wu2018understanding,gao2020patch,gao2020patch+,li2020regional,gao2021staircase,wang2021feature,fang2022learning,dong2018boosting,lin2019nesterov,zou2022making,wang2021boosting,Jang_2022_CVPR,wang2021enhancing,lu2021towards,he2022boosting,tan2022improving}
\newline
Better Loss: %
\cite{ganeshan2019fda,huang2019enhancing,inkawhich2020perturbing,inkawhich2020transferable, inkawhich2019feature,li2020towards,li2020yet,liu2019s, qin2021boosting,wang2021exploring,wang2020unified,wang2021feature, wu2020boosting,zhang2022improving,zhao2021success,zhou2018transferable,zhang2021data}\newline

Refine Surrogate: %
\cite{liu2016delving,li2020learning,zhu2021rethinking,zhou2022improving,wu2020skip,duan2022adversarial,guo2020backpropagating,zhang2021backpropagating,fang2022learning,yuan2021meta,wang2023rethinking,Xu_2023_ICCV}

& Improve UAP w/ Feedback: \cite{wu2020decision}\newline
Train Surrogate w/ Synthetic Data: %
\cite{papernot2016transferability,papernot2017practical,pengcheng2018query,zhou2020dast}\newline
Boost Existing Methods w/ Unlabeled Data~\cite{wang2020spanning}
& \emptysymbolcell & 
Boost Existing Methods:\newline
Trained Generator:
\cite{feng2022boosting,tu2019autozoom,mohaghegh2020advflow,bai2020improving,yatsura2021meta}, \newline Unlabeled Data~\cite{wang2020spanning}\newline
\textbf{w/ Pretrained Surrogate$^*$}:\newline
Save Queries with Surrogate:\newline
\cite{chen2021querynet,chen2022optimized,cheng2019improving,guo2019subspace,huang2019black,ma2020switching,ribeiro2016should,tashiro2020diversity,lord2022attacking,suya2020hybrid,yang2020learning}\newline

Refine Surrogate with Queries: \cite{cai2022blackbox,severi2022bad,yang2020learning}
\\ \cmidrule{2-6}

& Sufficient & 

Train Better (Deep) Surrogate:\newline
\cite{deng2021adversarial,salman2020adversarially,springer2021little,utrera2020adversarially, zhang2021early}\newline
Train AE Generator: %
\cite{baluja2017adversarial,bose2020adversarial,hashemi2020transferable,kanth2021learning, naseer2021generating,poursaeed2018generative}\newline
Input Transformation Network: %
\cite{yuan2021automa,wu2021improving,yuan2021adaptive}\newline
Train Simple Auxiliary Classifier:
\cite{inkawhich2020perturbing,inkawhich2020transferable,kanth2021learning}
& Improved Gradient Estimation w/ Trained Generator: \cite{zhang2021progressive,li2021nonlinear} & \emptysymbolcell & 
Train AE Generator:
\cite{yang2022boosting,du2019query,ma2021simulating,xiao2018generating}
\\ \bottomrule
\end{tabular}
}
\caption{Threat model taxonomy of black-box attacks.
The first two columns correspond to the quality and quantity of the auxiliary data available to the attacker initially. The remaining columns distinguish threat models based on the type of access they have to the target model, and for adversaries who can submit queries to the target model, the information they receive from the API in response. The symbol \emptysymbol\ above corresponds to areas in the threat-space that, to the best of our knowledge, are not considered by any attacks in the literature. The sub-category of \textit{w/ Pretrained Surrogate} with ``*" denotes that the corresponding attacks do not require auxiliary data, but the quality of data used to train the surrogate determines the corresponding cell. 
}
\label{tab:taxonomy}
\end{table*}

\subsection{Quantity of Initial Auxiliary Data}\label{sec:quant aux data}
Finally, we consider the quantity of auxiliary data (independent of data quality) \emph{initially} available to the adversary. We explicitly mention the availability of initial auxiliary data because the existence of some pretrained models may change the amount of auxiliary data available for the adversary in the end (\autoref{sec:pretrained models}). We consider two categories: the first is when the amount of data is only a handful and hence cannot be used to train models with decent performance for the attacks, \revision{while the second entails situations with enough data to train performant models.} Note that the definition of useful performance can vary depending on application scenarios, and we use this hypothetical and abstract description here. In practice, attackers may check whether the amount of available data can be used for training more useful models from the perspective of attack effectiveness (\eg the threshold can be set as the quantity sufficient to train a \revision{surrogate} classifier that is only X\% off compared to the \revision{prediction accuracy} of the target model). 

\begin{enumerate}[label=(\alph*)]
\item \textbf{Not Sufficient:} the quantity of data available is insufficient to train models useful for attacks. Attackers in this category may opt for leveraging other ways to utilize this information (\eg computing sample statistics \cite{wang2020spanning} or training shallow models \cite{li2020practical}). This category also contains the scenario of no auxiliary data (i.e., no samples). Strictly speaking, the ``quality" of the datasets does not matter as there is no auxiliary data at all, and this category falls ambiguously into any category of ``Auxiliary Data Quality". However, for clarity in presentation, we move attacks that do not require any auxiliary data into the category corresponding to the quality of ``No Overlap", to (best) denote that these attacks do not require any knowledge from the auxiliary data.

\item \textbf{Sufficient:} the quantity of data available is sufficient to train decent models (\eg generative models or classifiers), that can in turn assist with attacks.  
\end{enumerate}
\revision{
While attack strategies that require auxiliary data can technically be applied for any amount of data, implicit assumptions in such attacks may dictate certain requirements on data quantity for them to be effective. A discussion around the initial ``quantity'' of data is thus still relevant. 
For example, methods that require data to train well-performing surrogate models would understandably suffer from significant performance degradation when the amount of auxiliary data is limited, as demonstrated in ablation studies \cite{kanth2021learning}. However, the paper does not explicitly report the point at which attack performance drops to near-random. On the other hand, methods in ``Not Sufficient'' categories might face a bottleneck when given sufficient data, as the proposed approaches implicitly assume \emph{limited} data. Ablation studies on the impact of quantity of auxiliary data can be helpful to the community but are currently lacking in the literature. We advocate for including such studies in future works and discuss more in \autoref{sec:stronger_baselines}.
}
 
\subsection{Existence of Pretrained Models}\label{sec:pretrained models}
\revision{The literature has been historically building surrogate models directly from target models \cite{papernot2017practical}, and the availability of pretrained models today is an artifact of orthogonal advances in machine learning for building and releasing high-performing models, especially in the image domain. Such an assumption may not hold across other domains, especially in security-critical areas. We refer to such models as ``pretrained'' models. Assume a pretrained model trained on unknown proprietary data that is highly similar to the target model's data. An adversary that uses such a model implicitly leverages this data overlap through the publicly released model. To better capture this implicit leverage of high data quality, we classify attacks that only involve pretrained models into settings where the quantity of \emph{initial} auxiliary data is zero, and the quality of data is determined by the quality of private data used to train the model. For clarity, we add the existence of pretrained models as a sub-axis on top of the four main axes mentioned above.}

\section{Classification of Attacks on Threat Model}
\label{sec:attack category threat}

In this section, we categorize the black-box attacks based on their presumed threat model. \autoref{tab:taxonomy} presents our categorization of the surveyed attacks. 
The first main division is between attacks where the adversary has no interactive access to the target model, and ones where some level of interaction is available. Within each of these, we consider threat models based on the quality and quantity of data available to the adversary.
For the rest of this paper, we interchangeably use `transfer attacks' with `non-interactive attacks', and `query-based attacks' with `interactive attacks'.

\subsection{No Interactive Access to Target Model}

A significant fraction of attacks in the literature assume an adversary with no ability to submit queries and obtain responses from the target model. Without such access, the adversary has limited options and must use local resources to find good candidate examples.

\subsubsection{Low Quality Data: No Overlap with Target} %

This threat model assumes the least adversarial knowledge as the auxiliary data available for the attacker has no overlap with the training data for the target model, and the availability of the auxiliary data is limited. Works in this threat model have only appeared recently and to our knowledge, there is only one work that does not consider additional information (\eg pretrained models) and obtains successful adversarial examples with frequency manipulation \cite{zhang2022practical}.
A relaxation of this setting allows the adversary access to pretrained model(s) where the training set does not overlap with the target. As noted in \autoref{tab:taxonomy}, attacks in the literature that assume access to a pretrained surrogate do not leverage any additional auxiliary data and therefore, the quantity of auxiliary data is actually zero. 
Despite having a distribution mismatch,  surrogate model(s) can  capture some level of image semantics that can be valuable for  adversaries. Customized loss functions with respect to the pretrained models are designed by the adversaries to generate successful adversarial examples \cite{huan2020data,inkawhich2021can,kanth2021learning,lu2020enhancing,mopuri2018generalizable,mopuri2017fast,naseer2019cross,naseer2018task,qin2021adversarial,wu2020decision,zhang2021data,zhang2022beyond}. 
We note that some works \cite{inkawhich2021can,qin2021adversarial,zhang2022beyond,richards2021adversarial}  relax their setting to allow the auxiliary data to have partial overlap with the target in the data points and/or the labels, and also the availability of pretrained models (trained on data with partial overlap with target). These attacks still design customized loss functions to cope with distribution mismatch, and the (minor) difference to the ``no" overlap setting mainly lies in how to map the labels of the local surrogate to the labels of the target. As expected, attacks in partial-overlap settings achieve better results than ones with no overlap. To the best of out knowledge, no work in the literature assumes sufficient low-quality (no/partial overlap with target) auxiliary data, while this situation is likely to be common in practice. For example, when attacking some unknown target model (\eg medical image classifier \cite{dong2023adversarial}), attackers may leverage the ImageNet dataset.

\subsubsection{High Quality Data: Complete Overlap with Target}
The distribution of auxiliary data is highly similar (or even the same) to the target training distribution. Under limited availability of such data, shallow surrogates can be trained to enable higher transferability \cite{li2020practical,sun2022towards}. This assumption may be further relaxed when adversaries have access to some pretrained models trained on high-quality auxiliary data. Like the case of low-quality auxiliary data, existing works that use pretrained models do not utilize auxiliary data. This is the most explored attack setting in the literature: methods include gradient stabilization ~\cite{wang2022enhancing,wu2018understanding,gao2020patch,gao2020patch+,li2020regional,gao2021staircase,wang2021feature,fang2022learning,dong2018boosting,lin2019nesterov,zou2022making,wang2021boosting,Jang_2022_CVPR,wang2021enhancing,lu2021towards,he2022boosting,tan2022improving}, input augmentation \cite{dong2019evading,Jang_2022_CVPR,lin2019nesterov,long2022frequency,wang2021admix,wu2021improving,xie2019improving,yang2022adversarial,yuan2021adaptive,yuan2021automa,zou2020improving,Byun_2022_CVPR,wang2023structure,zhang2023improving,fang2022learning}, better loss designs \cite{ganeshan2019fda,huang2019enhancing,inkawhich2020perturbing,inkawhich2020transferable, inkawhich2019feature,li2020towards,li2020yet,liu2019s, qin2021boosting,wang2021exploring,wang2020unified,wang2021feature, wu2020boosting,zhang2022improving,zhao2021success,zhou2018transferable,zhang2021data} and surrogate refinement \cite{liu2016delving,li2020learning,salman2020adversarially,utrera2020adversarially,deng2021adversarial,zhou2022improving,wu2020skip,duan2022adversarial,yuan2021meta,fang2022learning,zhu2021rethinking,zhang2021backpropagating,guo2020backpropagating,springer2021little,zhang2021early,yang2022boosting}, as discussed in \autoref{sec:transfer attacks}. \revision{One example of a scenario with an insufficient amount of high-quality auxiliary data is the case of a face recognition target model. In this context, auxiliary data might only consist of a few face images captured under the same conditions (such as the same setting, background, etc.) as the target model's training data, but acquiring a large amount of such high-quality data can be challenging.}

When there are sufficient amount of high-quality auxiliary data available (attackers can also naturally obtain well-performing surrogate models), the proposed methods can be quite diverse: directly training better surrogate classifiers to generate more transferable adversarial examples \cite{deng2021adversarial,salman2020adversarially,springer2021little,utrera2020adversarially, zhang2021early}, training auxiliary classifiers on top of the surrogate classifiers \cite{inkawhich2020perturbing,inkawhich2020transferable,kanth2021learning}, training generators to generate likely-to-transfer adversarial examples \cite{baluja2017adversarial,bose2020adversarial,hashemi2020transferable,kanth2021learning, naseer2021generating,poursaeed2018generative}. Besides these methods, some attacks also focus on finding better transformation methods with neural networks \cite{yuan2021automa,wu2021improving,yuan2021adaptive} so that these inputs, when input to some surrogate classifiers, can lead to improved transferability. Notably, many of these attacks (\eg training auxiliary classifiers and finding better input transformations) are compatible with each other, indicating that stronger attacks might be possible by composing these attacks, which are not explored in the literature, and we encourage researchers to investigate this possibility. %

\subsection{Hard-Label with Interactive Access}

In this subsection, we consider attacks where the adversary can actively query the target model, but only receives hard-label responses. Within this category, we break down attacks according to the auxiliary data available to the adversary, following a structure similar to that of the previous subsection. %

\subsubsection{Low Quality Data: No Overlap with Target}
Attacks in this category are rather restricted in terms of the attacker knowledge as existing attacks in the literature in fact did not utilize any auxiliary data, leading to the category of the quantity of auxiliary data being zero. The quality of data should not be relevant in this case, but we still put it into the setting of ``no overlap with target" mainly for convenience in categorization. Despite being a challenging setting, many hard-label query-based attacks are proposed. 
The common methods include estimating the gradients \cite{cheng2018query,cheng2019sign,chen2020hopskipjumpattack,liu2019geometry,simon2021popskipjump,wang2022query,rahmati2020geoda,zhao2021improved}, deploying some classic black-box optimization techniques \cite{shukla2021simple,zhao2019design}, leveraging random-walk strategy  \cite{brendel2017decision,brunner2019guessing,dong2019efficient,li2021aha,shi2020polishing,sun2022query,maho2021surfree}, or developing other gradient-free random search based methods \cite{chen2020rays,chen2020boosting,midtlid2022magnitude,vo2022query,wang2022triangle}. The categories that allow adversaries to leverage some (sufficient or insufficient but not zero) amount of auxiliary data are largely missing from the current literature.

\subsection{High Quality Data: Complete Overlap with Target}
Attacks in this category have access to auxiliary data sampled from a distribution highly similar to/same as the target distribution. When the amount of auxiliary data is insufficient, the proposed methods include finding better (untargeted) universal adversarial perturbations that are agnostic to the victim images \cite{wu2020decision}, training surrogate models using synthetic dataset \cite{papernot2016transferability,papernot2017practical,pengcheng2018query,zhou2020dast} and boost existing hard-label attacks using limited amounts of unlabeled dataset~\cite{wang2020spanning}. When sufficient auxiliary data is available, this data can be used to train generators to obtain better gradient estimates \cite{zhang2021progressive,li2021nonlinear}. Interestingly, the number of works published under this category is still much less compared to the more restrictive category above.

\subsection{Top-K Confidence Vector with Interactive Access}
Attacks in this category can interact with the target model and get the \topk\ part of the confidence vector from the target model. So far, there is only one work \cite{ilyas2018black} that explicitly designs an attack for this setting, although such a scenario is also very common in practice. 
Driven by limited exploration in this category, we conduct preliminary experiments in \autoref{sec:unexplored areas} to show that, currently under-explored areas may not be solved by trivially adapting techniques from other well-explored areas and motivate future investigation along this direction.

\subsection{Complete Confidence Vector with Interactive Access}
Attacks in this category will receive the complete prediction confidence vector returned from the target model. The remaining breakdowns are still similarly based on the quality and quantity of the auxiliary data available.

\subsubsection{Low Quality Data: No Overlap with Target}
The strictest setting is when the adversaries do not use any auxiliary data. In this setting, many works propose generating highly successful adversarial examples (\eg finding many untargeted adversarial examples in $<100$ queries). Typical methods include gradient estimation\cite{chen2017zoo,al2019sign,chen2020frank,bhagoji2018practical,du2018towards, ilyas2018black,ilyas2018prior,li2020projection,li2019nattack, liu2018signsgd,shi2019curls,uesato2018adversarial,zhao2020towards, zhao2019design}, leveraging classical black-box optimization techniques \cite{chen2019poba,alzantot2019genattack,meunier2019yet,ru2019bayesopt,suya2017query} or proposing some efficient random search methods \cite{chen2019poba,alzantot2019genattack,croce2022sparse,croce2019sparse,guo2019simple,moon2019parsimonious,meunier2019yet,narodytska2016simple,tran2022exploiting,shiva2017simple,andriushchenko2020square}. 

When the assumption is relaxed to allow limited number of auxiliary data that overlaps with the target distribution partially, a generator \cite{yatsura2021meta} on the perturbation distribution can be trained to boost the performance of the state-of-the-art Square Attack \cite{andriushchenko2020square} that does not use any auxiliary data. 

\subsubsection{High Quality Data: Complete Overlap with Target}
Under limited availability of high-quality auxiliary data, existing works train generators to improve performance by better capturing the low-dimensional latent space where the adversarial examples reside \cite{feng2022boosting,tu2019autozoom,mohaghegh2020advflow,bai2020improving,yatsura2021meta}. The availability of some (auxiliary) unlabeled data also improves existing attacks that (originally) do not rely on auxiliary data \cite{wang2020spanning}. Like the low-quality data case, a generator for the perturbation distribution can still be trained on the limited high-quality auxiliary data to boost performance of the Square Attack \cite{yatsura2021meta}. When the assumption is further relaxed to allow some pretrained models trained on data highly similar to the target's training data, the pretrained models can be used to boost query-based attacks \cite{chen2021querynet,chen2022optimized,cheng2019improving,guo2019subspace,huang2019black,ma2020switching,ribeiro2016should,tashiro2020diversity,lord2022attacking,suya2020hybrid,yang2020learning} or queries from the target model can be used to refine the surrogate model \cite{cai2022blackbox,severi2022bad,yang2020learning}. The most relaxed setting is when there are sufficient high-quality auxiliary data available. Existing works train generators on (sufficient) high-quality data to generate adversarial examples directly. In particular, a generator is first trained on some local surrogate models (can be easily obtained by training on the auxiliary data if not available beforehand) and later fine-tuned with queries from the unknown target model \cite{yang2022boosting,du2019query,ma2021simulating,xiao2018generating}.

\section{Insights from Taxonomy} \label{sec:taxonomy_insights}

Studying published attacks from the perspective of our threat model taxonomy results in several insights about gaps in the current research 
(\autoref{sec:unexplored areas}), ways to improve evaluation (\autoref{sec:stronger_baselines}), and opportunities to improve techniques by incorporating ideas from related fields, such as model extraction and inversion (\autoref{sec:new direction}).

\subsection{Technical challenges in Underexplored Areas}
\label{sec:unexplored areas}

As can be seen from \autoref{tab:taxonomy}, many threat models are unexplored (marked with \emptysymbol) or have only been considered by a few works. Across the rows, there is little work in settings where ample data is available but from sources that have limited overlap with the target model's data distribution. However, this is perhaps the most relevant practical scenario---for most classification tasks, adversaries are likely to be able to acquire large amounts of somewhat similar data (\eg from the Internet, open image datasets), but unlikely to be able to sample from the same distribution as the target model's (private) training distribution. 

Across the columns, only one attack explicitly optimizes for the availability of \topk\ prediction scores. This is surprising since this is the most likely scenario for API attacks on deployed classifiers. For example, ClarifAI's models\footnote{\url{https://docs.clarifai.com/api-guide/predict/prediction-parameters/}} return scores for at most 200 classes.
For these unexplored or under-explored settings, we suspect there is a technical gap in addition to a knowledge gap, so the settings cannot be addressed satisfactorily by adapting state-of-the-art methods from well-explored areas \cite{zhang2022beyond,richards2021adversarial}. To support our argument, we propose an attack for the \topk\ setting, specifically for the setting with no auxiliary data or pretrained models is available, the typical setting for query-based attacks (\autoref{sec:query attacks}). Our adapted attack is based on the Square Attack~\cite{andriushchenko2020square} that is originally designed for the setting that receives full confidence vector of prediction and the adaptation idea is built on top of the design of NES: \topk\ attack in Ilyas et al.~\cite{ilyas2018black} with non-trivial modifications (details in \Cref{app:topk adaptation}).

\begin{figure}[tbh]
    \centering
    \includegraphics[width=0.9\linewidth]{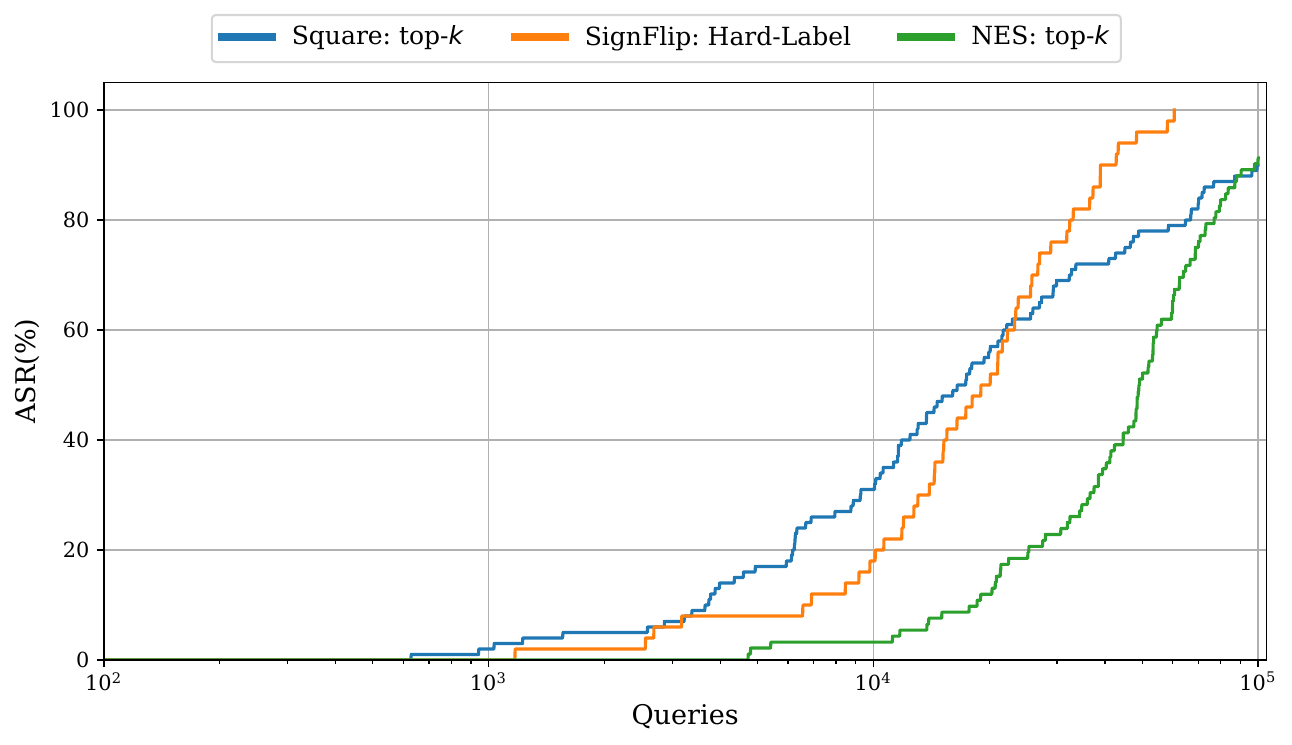}
    \caption{Comparison of \topk\ attacks. \topksquare\ is our proposed adaption of the Square Attack for the \topk\ setting. NES: \topk\ is the current state-of-the-art attack. SignFlip \cite{chen2020boosting} is a more restrictive hard-label attack.}
    \label{fig:topk comparison}
\end{figure}

Following the setup in the baseline NES: \topk\ \cite{ilyas2018black}, we consider targeted attacks, set the query limit to 100,000 for both attacks, and assume only the top-$1$ prediction confidence is available. As shown in \autoref{fig:topk comparison}, the \topksquare\ outperforms the NES: \topk\ attack significantly as the number of queries increases. However, the \topksquare\ is mostly outperformed by the hard-label SignFlipAttack \cite{chen2020boosting}, which ensures the targeted label $y_t$ is always in the top-$1$ prediction and then chooses to ignore the extra prediction confidence. This comparison illustrates that there is substantial room for improving attacks in the \topk\ setting, as attacks designed for this setting are not performing nearly as well as attacks with less information. Moreover, it is essential to underscore the significance of considering baseline attacks that operate with a subset of the available information in the given context. One might improve further \topk\ attack performance by harnessing the available confidence score for hard-label attacks or adapting techniques from multi-label learning \cite{Hu_2021_ICCV}. However, this task is not straightforward, as our preliminary experiments suggest. \revision{In general, research in the underexplored areas, such as the ones we outline, faces two unique challenges. First, well-explored methods that require extra information do not directly extend to other threat models, and their adaptation can be complex, with no reasonable estimates on the performance drops---we demonstrate one such case in our experiments in \Cref{fig:topk comparison}. Second, well-explored attacks from more restrictive settings (\ie less information) can be trivially extended to less restrictive settings, but the room for improvement with additional information provided by such less restrictive settings is unclear. For instance, the state-of-the-art hard-label attack has a success rate of 20\% at 10,000 query limit, while the full-score setting has near-perfect attack success at the same number of queries. While one can argue that hard-label attacks can be used in \topk\ setting and yield 20\% success rate, it is intuitively clear that additional information via \topk\ should be usable to increase success rates significantly. While our adaptation (\Cref{fig:topk comparison}) achieves a success rate of over 30\%, there may still be room for improvement. We thus encourage researchers to concentrate on crafting and examining attacks designed for relevant threat model scenarios, such as the ones we identify.}

\subsection{Stronger Baselines Under Same Threat Model}
\label{sec:stronger_baselines}

\begin{table}[t]
    \centering
    \begin{tabular}{cccc}
         \toprule
         Attacks & Square Attack & \odsrgf & \squarenew\\ \midrule
         Attack Success (\%) & 100 & 97.7 & 100 \\
         Average Queries & 2,317 & 1,242 & 117\\
         \bottomrule
    \end{tabular}
    \caption{Comparing query-based attacks in a setting where all attacks are given access to an ensemble of four surrogate models. 
Experimental setup follows from Tashiro et al.~\cite{tashiro2020diversity}. \squarenew\ is our proposed stronger baseline.
}
    \label{tab:stronger baseline}
\end{table}

Works introducing black-box attacks often make various assumptions about the knowledge of the adversary and often end up comparing adversaries across different levels of knowledge directly in terms of attack effectiveness. We advocate that with the categorization of the threat space (as outlined in~\Cref{tab:taxonomy}), attacks should be carefully compared within the same threat space. Further, researchers should be mindful of the possibility of combining additional information made available to the adversary to design stronger baselines. 

Here, we use a preliminary experiment on the category of complete access to prediction vectors and an ensemble of local surrogates to demonstrate that, when evaluated under the same threat model, a strong baseline can exist (and be easily found) to overturn the state-of-the-art claims in the paper. Specifically, \odsrgf\ \cite{tashiro2020diversity} leverages diversified gradient vectors from the local surrogate models as the perturbation vector for the RGF attack~\cite{cheng2019improving}. This attack performs better than the Square Attack~\cite{andriushchenko2020square} that does not require any pretrained surrogate models. Using a simple strategy of generating candidate adversarial examples against the (assumed) local surrogates, followed by running the Square Attack on the remaining examples that fail to transfer from, can easily establish a (much) stronger baseline. This idea is inspired by Suya et al.~\cite{suya2020hybrid}, which appears before the \odsrgf\ attack \cite{tashiro2020diversity}. Details of the transfer experiment (on generating local adversarial examples) can be found in \Cref{app:implementation}. \autoref{tab:stronger baseline} compares \odsrgf, Square Attack and our proposed \squarenew\ in terms of the attack success rate and the average number of queries, using the same experimental setup as the original apper \cite{tashiro2020diversity}. The first two attacks are the proposed and baselinesattacks in Tashiro et al.~\cite{tashiro2020diversity}. We observe that both \odsrgf\ and \squarenew\ improve query efficiency compared to the original Square Attack. However, the \squarenew\ attack significantly outperforms the proposed \odsrgf\ attack, demonstrating the importance of considering simple adaptations of known attacks to new threat models.

\revision{At last, stronger baselines may emerge not only when extra information is available but also when attacks utilize auxiliary data, even in the absence of such extra information. As mentioned in \autoref{sec:quant aux data}, it is worth noting that attacks that operate with auxiliary data can theoretically be applied in settings with varying data sizes. The key distinction lies in the degree of effectiveness these attacks exhibit under different data sizes. Therefore, we recommend that attack methods, which implicitly assume the availability of ``sufficient'' or ``insufficient'' auxiliary data, should also use methods from the opposite category as baselines. Furthermore, researchers should conduct ablation studies to examine how the attack performance evolves, compared to the baselines, when transitioning from ``insufficient'' to ``sufficient'' auxiliary data.}

\subsection{Interaction Among Attacker Knowledge}
\label{sec:new direction}

The most straightforward interaction of attacker knowledge is adversaries can train many pretrained models given enough auxiliary data. Therefore, attacks may treat the existence of sufficient auxiliary data the same as the existence of both the data and the pretrained models (obtained from the data). Further, proper identification of threat models using our taxonomy uncovers connections to other related fields \revision{such as model stealing (also known as model extraction) \cite{tramer2016stealing} and model inversion \cite{fredrikson2015model}. Model stealing adversaries aim to steal a copy of a remotely deployed machine learning model given Oracle prediction access. In contrast, model inversion adversaries seek to infer (parts of) the training distribution of the remote model. These attacks}
can significantly boost the performance of black-box attacks with interactive access to the target model by providing better surrogate models (via model extraction) and more representative training data (via model inversion). We do not implement these ideas but discuss their potential in detail below.

\shortsection{Model-Extraction Attacks}
Simply identifying the target model structure (or family of models)~\cite{oh2019towards,jagielski2020high} can improve attack success, especially in settings where the auxiliary data highly overlaps with the target model's training data. The extensive literature on attack transferability~\cite{dong2018boosting,xie2019improving,inkawhich2019feature,inkawhich2020transferable} can thus serve as a ``handbook" for adversaries.
Further, when we look to utilize model extraction for better transferability of adversarial examples, an adversary's specific goal is to ensure the extracted and victim models have a similar vulnerability space, so that better surrogates can boost black-box attack performance. This is an easier objective that the original model extraction objective of having prediction consistency~\cite{oliynyk2023know} as it is believed that adversarial examples reside in a low dimensional subspace~\cite{jagielski2020high} that is easier to capture than the full input space.

When enough auxiliary data exists, state-of-the-art model extraction attacks can be readily applied~\cite{jagielski2020high}. Limited auxiliary data settings are more challenging. Several works on black-box adversarial examples use surrogate training to enhance transferability~\cite{papernot2016transferability,papernot2017practical,pengcheng2018query} or improve query efficiency~\cite{yang2020learning}. Surrogate training is also common in model extraction \cite{oliynyk2023know}. However, these surrogate extraction methods fail for complex image classification tasks. Recent advances in data-free extraction attacks show promise for addressing complex classification tasks and can be further enhanced with pretrained models.

Data-free model-extraction attacks~\cite{kariyappa2021maze,truong2021data,rosenthal2023disguide,sanyal2022towards} rely on a generator to generate queries, which are then labeled by the target model and used to update the generator and the extracted model. 
These methods work well without any pre-trained model---in particular, generators are randomly initialized and optimized with the estimated gradient from the target model~\cite{kariyappa2021maze,truong2021data}. 
With pretrained models, one may first (pre-)train a generator with auxiliary models (using their actual gradients) and then continue training the generator with estimated gradients from the black-box model. Such a generator is likely better than a randomly initialized one and may enable extraction in fewer queries. The feasibility of pretraining a generator and then fine-tuning for the target model has already been demonstrated when directly generating adversarial examples~\cite{yang2022boosting,du2019query,ma2021simulating,xiao2018generating}. Additionally, knowledge from the surrogate models may still transfer to the target when the training data of the two models have partial or no overlap \cite{yatsura2021meta}. We note that the obtained generator can also be used to augment the adversary's data. When limited quantities of data are available, this increased data can in turn enable other model extraction methods that require more auxiliary data~\cite{jagielski2020high}.

\shortsection{Model-Inversion Attacks}
Model-inversion attacks aim to recover representative and semantically meaningful training data~\cite{chen2021knowledge} of a given model. However, to generate adversarial examples, the extracted data need not be semantically meaningful~\cite{yang2022boosting,zhang2021progressive}. 
Model inversion can help either directly~\cite{yang2022boosting,du2019query,ma2021simulating,xiao2018generating,zhang2021progressive,li2021nonlinear} by providing more representative data (which can be further diversified with data augmentations~\cite{gong2021inversenet}), or indirectly by boosting the performance of model extraction attacks via better query synthesis~\cite{gong2021inversenet}. 
In settings with sufficient auxiliary data, state-of-the-art model-inversion attacks~\cite{yuan2023pseudo,liu2023unstoppable} can be applied directly to recover more representative data and improve the quality of the auxiliary data.
For settings with limited auxiliary data, an adversary may use a query-generator trained during a model-extraction attack (where the generator is a common component in most techniques, as described earlier) to generate more auxiliary data. State-of-the-art black-box inversion attacks~\cite{liu2023unstoppable} can then be utilized in the absence of a surrogate model.

Still motivated by the success of pretraining and fine-tuning generators for adversarial example generation~\cite{yang2022boosting,du2019query,ma2021simulating,xiao2018generating}, we see opportunities for exploiting the presence of pretrained auxiliary models in improving the effectiveness of model inversion attacks against the unknown target, to eventually improve performance of black-box attacks. 
Particularly, the conditional generative model in Liu et al.~\cite{liu2023unstoppable} can first use labels from auxiliary models, followed by fine-tuning with labels from the target model to improve performance. Similarly, white-box inversion attacks~\cite{yuan2023pseudo} may utilize the auxiliary model for gradient computation and then use predictions from the target model to estimate gradients using black-box gradient estimation~\cite{chen2017zoo,ilyas2018black} techniques for fine-tuning.

\shortsection{Combining Model-Stealing and Model-Inversion}
Model stealing and model inversion attacks can be combined dynamically---for instance, by iteratively running model stealing and inversion attacks to boost each other. One thing to note is that these attacks' query requirements can be quite high and unrealistic for resource-constrained adversaries, even though attackers only have to run these two attacks once and then use the results to boost future black-box attacks. 
For example, even state-of-the-art black-box model inversion attacks require millions of queries (\eg DiSGUIDE~\cite{rosenthal2023disguide} use at least 4M queries for models trained on CIFAR-10 \cite{krizhevsky2009learning}).

\section{Rethinking Baseline Comparisons}
\label{sec:new_eval}

\begin{figure*}
    \includegraphics[width=\figscale\textwidth]{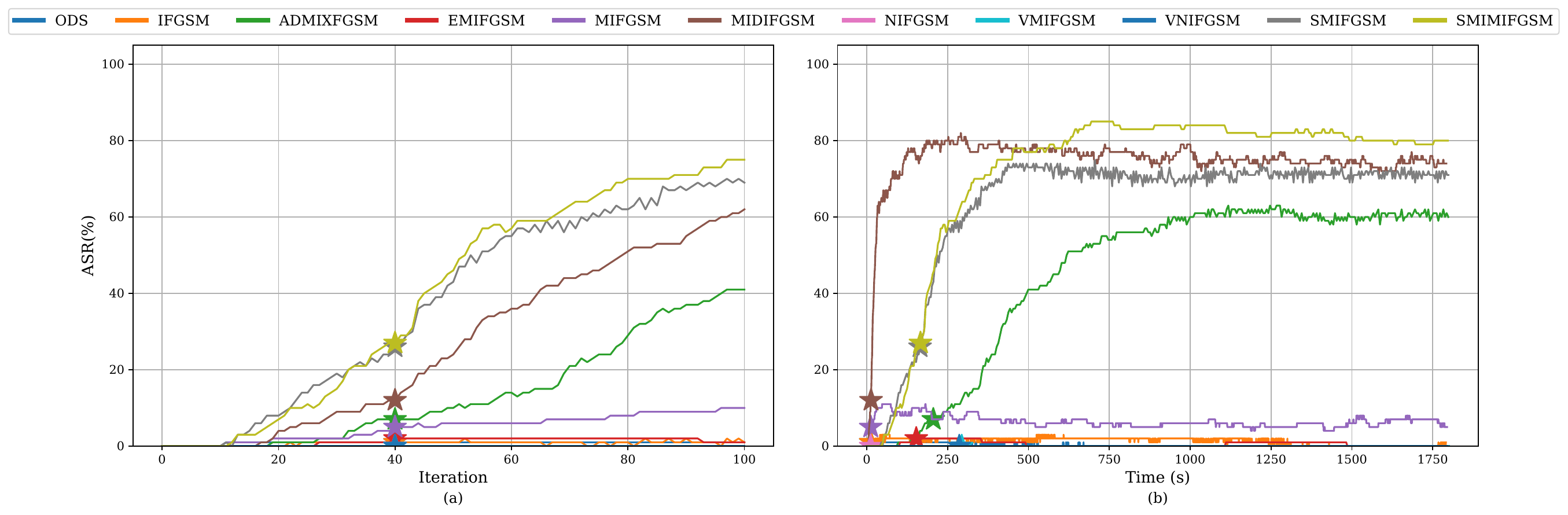}
    \caption{ASR (y-axis) for various targeted attacks on DenseNet201 models, varying across iterations (a) and time (b). All attacks on the left are run for 100 iterations, while attacks on the right are run for 30 minutes per batch. ASR at each iteration is computed using adversarial examples at that iteration. ASR at \niterstargeted iterations are marked with $\star$ for each attack.}
    \label{fig:time_vs_iters}
\end{figure*}

Most interactive and non-interactive attacks involve running an optimization loop locally for some number of iterations to find a candidate adversarial example. It is in the adversary's interest to run the attack for as many iterations as possible as long as more iterations improve success rates. The number of iterations is also used as a grounding factor in attack comparisons, running attacks for the same number of iterations for fair comparison~\cite{wang2021admix, dong2019evading,wang2021boosting,wang2022enhancing,wang2021enhancing,dong2018boosting}.
\footnote{Ablation studies report the impact of iteration numbers on ASR, but only up to 30 iterations \cite{xie2019improving}. Studies on higher attack iterations (up to 1,000) do not report iteration-wise results \cite{richards2021adversarial}.
}
We argue that such measures, while well-meaning, are in fact not ``fair" and misaligned with what adversaries care about. Fixing the number of iterations limits some attacks, clipping their potential for the sake of comparison. In most cases, the iteration-wise cost of attacks is low, and an adversary that does not have severe latency requirements would only care about maximizing its success rate. When latency or compute costs matter, an adversary would prefer the attack that yields the highest attack success rate within the given time or resource constraints.
As pointed out by Apruzzese et al.~\cite{apruzzese2023real} through a thorough analysis of real-world adversarial scenarios, attackers prefer cheap and effective methods \revision{that can be easily automated}, and the relevant cost metric is \revision{the total effort spent on the process of completing an attack}---a metric that is harder to count than number of iterations, but is more direct. 

Another issue with many evaluations is the lack of challenging settings for attack comparisons. Untargeted attacks are much easier than targeted ones; attacking non-robust standard models is easier than attacking adversarially-robust models. Success rates can be very high in easy settings and thus fail to provide useful insights about relative attack performance that can transfer to harder settings. These harder settings are in fact the ones where attacks matter most.

We advocate for evaluating black-box attacks with a realistic consideration of actions that adversaries can take in practice. Specifically, instead of fixating on a specific number of iterations for non-interactive settings (\autoref{sec:runtime_transfer_exps}) as the primary metric for comparison, we argue that adversaries should be able to use more iterations when beneficial and the only constraints such as total time should be motivated by the evaluation scenario. Our analysis with this new lens uncovers several interesting insights and suggestions for researchers. For non-interactive transfer attacks, we discover how running attacks for more iterations helps attack success (\Cref{sec:time_vs_iters}), and that simply stopping attacks when they succeed on local models can hamper performance (\Cref{sec:more_iters}), and observe much clearer trends in relative attack performance trends when evaluated in hard settings such as targeted attacks (\Cref{sec:harder_settings}). In \Cref{sec:runtime_query_exps}, we also show how the ASR of different interactive query-attacks can change when the evaluation metric shifts from the number of queries to the local runtime\revision{, and advocate using local runtime as an additional metric on top of the commonly considered query costs for the interactive black-box attacks}.

\subsection{Transfer Attacks} \label{sec:runtime_transfer_exps}
Attack success rate (ASR) has been the guiding metric for evaluating different transfer attacks' the effectiveness of different transfer attacks. However, more effective transfer attacks often require complicated computation processes and can lead to local computation costs that are orders of magnitude higher than baselines.

For convenience in comparison, we selected transfer attacks that augment the baseline \ifgsm\ attack \cite{kurakin2016adversarial} with various gradient and input manipulation techniques, including new combinations (details in \Cref{sec:evaluated attack background})---this leaves us with 20 attacks. 
Since these attacks are based on iterative local optimizations, we can conveniently measure the impact of different local time constraints on ASR against the target models. Of these 20 attacks, we picked 11 that span a wide range of local runtimes.
Note that for each of the following graphs, we re-evaluate the attack at each iteration using the adversarial inputs generated at the end of that iteration, thus giving us multiple attack success rates as iterations progress.

\subsubsection{Time and iterations} \label{sec:time_vs_iters}
It is conventional to evaluate attacks for a fixed number of iterations: usually 10 for untargeted settings. However, the lack of targeted attack evaluations means there is no such standard for that setting, with \mifgsm~\cite{dong2018boosting} being one of the few attacks that evaluate targeted attacks, using \niterstargeted iterations (which is what we set for targeted attacks). However, the number is arbitrary and it is unclear whether attacks have the potential to have improved performance. Prior work\cite{carlini2019evaluating} also recommends running attacks until convergence, instead of a fixed number of iterations.
To test this hypothesis, we run attacks for 100 iterations, instead of the usual \niterstargeted for targeted attacks, and analyze ASR trends (\autoref{fig:time_vs_iters}-a). Most attacks seem to benefit from increased iterations.

Given this potential for improved success beyond \niterstargeted iterations, it is important to extend evaluations for valid comparisons. Execution time should be used if resource constraints like runtime exist, especially when the cost-per-iteration varies. Motivated by these factors, we re-run all the attacks but instead of running them for a fixed number of iterations as in prior work, we run them for the same time duration (30 minutes per batch)\footnote{We opt for measuring total runtime over algorithmic measures due to the challenge in standardizing components' runtimes across different hardware configurations, acknowledging both methods have their merits and limitations.}.

Iteration-wise analysis (\autoref{fig:time_vs_iters}-a) would suggest \midifgsm\ to be slightly worse off than \smifgsm, even when compared under the setting of 100 iterations. However, looking at the same results across time (\autoref{fig:time_vs_iters}-b), this trend flips once we observe that \midifgsm\ is nearly 2x as fast and can thus execute double the number of iterations in the same amount of time. Similarly, \midifgsm\ and \admixfgsm\ do not seem very far apart in their performance when looking at the same number of iterations, but time-wise analysis shows how the difference in their performance is much higher.

This analysis based on runtimes paints a clearer picture that is better aligned with what an adversary would desire---maximizing attack success within their available \revision{resources (e.g., limit on the total execution time)}.
\revision{As an example, consider an adversary looking at results in the literature to select an attack. Assuming computational constraints are not an issue, the literature would suggest \smifgsm\ being a good candidate instead of candidates like \midifgsm, and the adversary may pick \smifgsm\ to conduct the attack. However, once we realize that these comparisons are based on the number of attack iterations (an arbitrary metric) and instead compare them based on the local runtime, it is clear that \midifgsm\ can generate better attack results (\Cref{fig:time_vs_iters}).
These are the kind of cases we have in mind while advocating for time-based comparisons. Our motive is not to encourage researchers to add execution-time as an "extra metric", but rather remember that these attacks are designed for adversaries that would only want to maximize success rates given available resources \cite{apruzzese2023real}, and not care about running the attack for a fixed number of attack iterations.}
\begin{mdframed}
\textbf{Recommendation}: Run attacks for enough iterations until attack success rates plateau. Execution cost such as the local attack runtime should be used as the equalizing factor when comparing black-box attack performance, not the number of iterations.
\end{mdframed}

\begin{figure}
    \includegraphics[width=\linewidth]{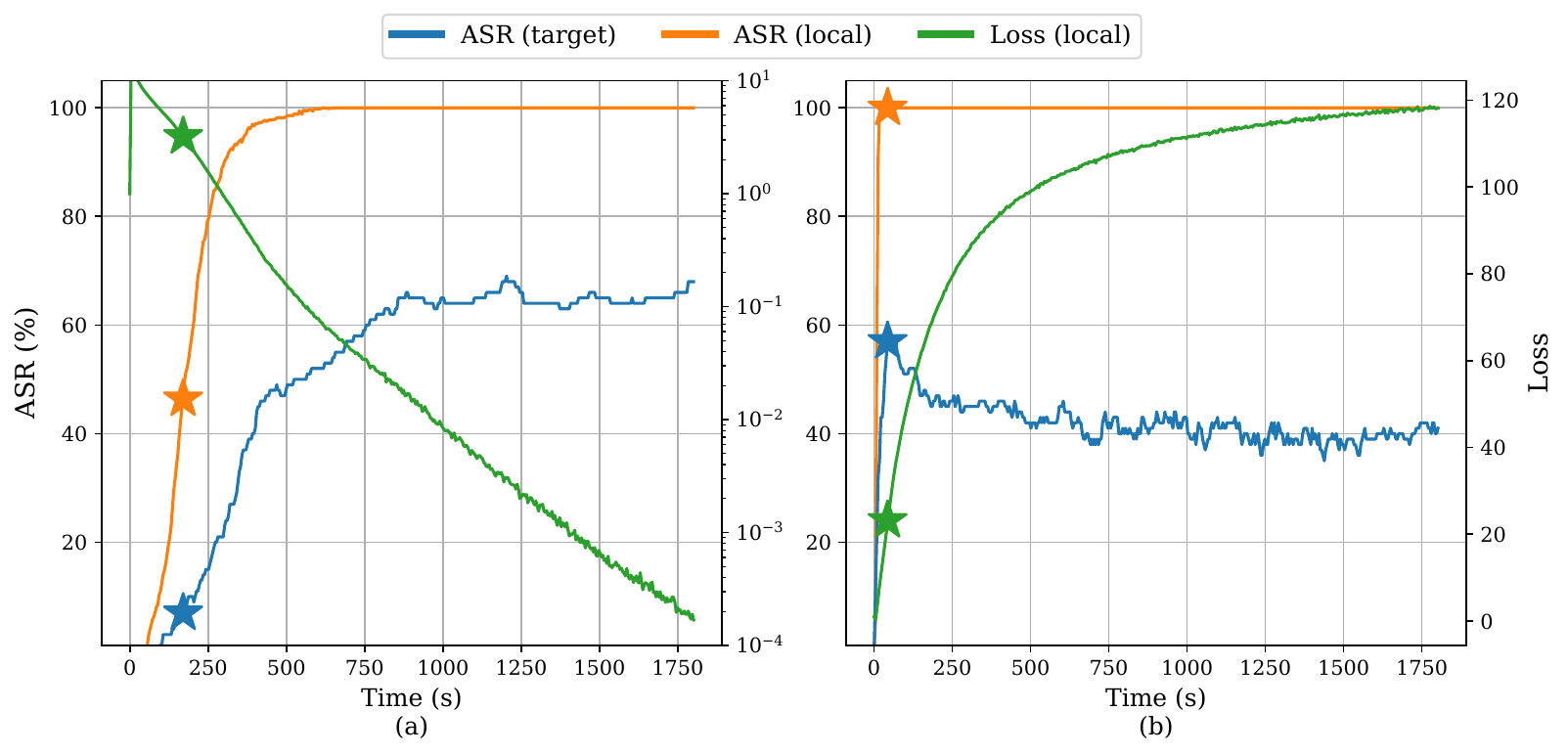}
    \caption{Attack success rates (ASR) (y-axis, left) for target and local models, along with loss (y-axis, right) while optimizing the objective locally, varying across time (x-axis), for targeted attacks on DenseNet201 (a) and untargeted attacks on adversarially-robust Inception-v3$_\text{adv}$ (b), using \smimifgsm\ \cite{wang2022enhancing}.
    ASR at representative iterations (\niterstargeted for targeted, 10 for untargeted) are marked with a $\star$ for each of the metrics.}
    \label{fig:local_metrics_track}
\end{figure}

\subsubsection{Knowing When to Stop} \label{sec:more_iters}
As observed in \autoref{fig:time_vs_iters}, simply running attacks for more iterations often improves attack success rates. For instance, attack success for \midifgsm\ jumps from $<20\%$ to nearly $80\%$ when run for sufficient iterations which, interestingly, is still faster than running \admixfgsm\ for \niterstargeted iterations. Similarly, \smimifgsm\ jumps from ${\sim}75\%$ to ${\sim}85\%$, once the attack runs for longer.
However, success rates do not always improve with more iterations. For instance, while \midifgsm\ in the targeted setting (\autoref{fig:time_vs_iters}-b) sees an improvement, it fluctuates between $70$ and $80\%$.
While running attacks for more iterations helps in most cases, it is not obvious when an adversary should stop their attack to maximize ASR---the adversary cannot know the optimal number of iterations before executing their attack. One possible workaround is keeping track of metrics for the local models (which are used to compute gradients), and possibly running more iterations as long as metrics such as local success rates and loss do not stagnate.

\begin{figure*}[t]
    \includegraphics[width=\figscale\textwidth]{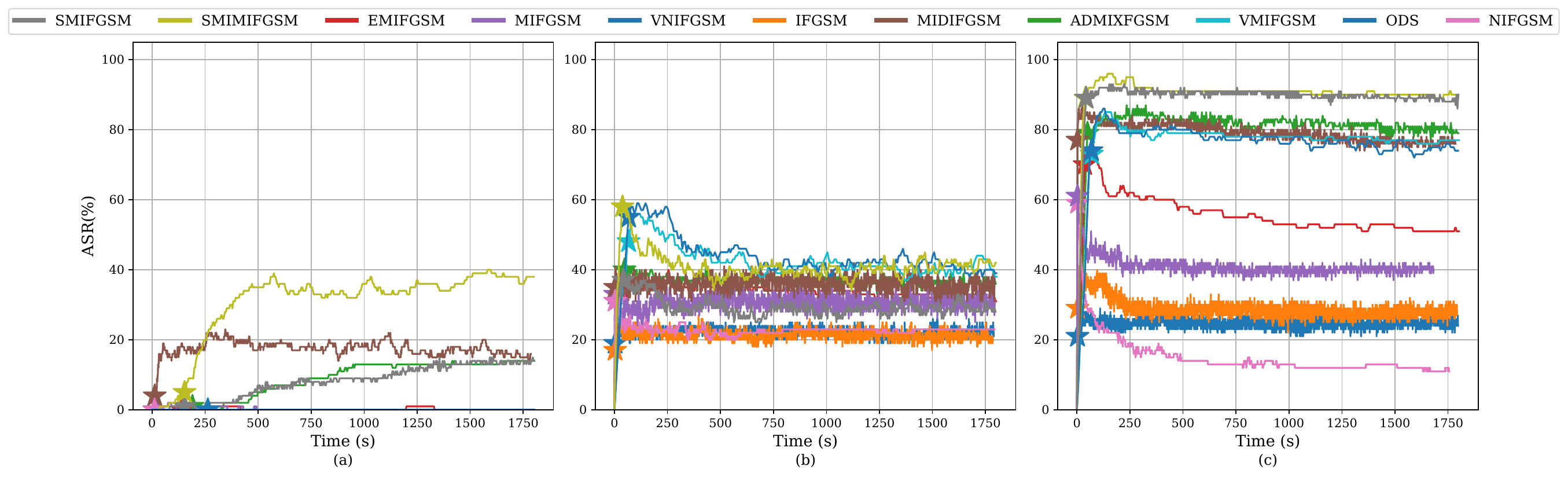}
    \caption{ASR (y-axis) for various attacks: targeted attacks for Inception-v3 with perturbation budget $16/255$ ($\ell_\infty$) (a), untargeted attacks for Inception-v3 with reduced perturbation budget $8/255$ (b), and untargeted attacks for adversarially robust model Inc-v3$_{\text{adv}}$ with perturbation budget $16/255$ (c). ASR at each iteration is computed using adversarial examples at that iteration. ASR at representative iterations (\niterstargeted for targeted, 10 for untargeted) are marked with $\star$ for each attack.}
    \label{fig:eps16_vs_eps8_vs_adv}
\end{figure*}

Intuitively, an adversary has no reason to continue local attack optimization once it successfully generates adversarial examples for its local models. The only possible motivation lies in changing the model's prediction probabilities---increasing confidence for targeted attacks, and decreasing confidence for untargeted attacks.
Our analysis shows how the rate of finding successful adversarial examples against local models gets to 100\% almost immediately, even when attack success rates on the target models are low. An adversary that only inspects local attack success rates would thus stop its optimization prematurely, leading to sub-optimal ASR for the target model.

For the targeted setting (\autoref{fig:local_metrics_track}-a), we interestingly observe the local loss value to continue dropping (not by much; note that the right y-axis is on log-scale for loss in targeted attack), even though target ASR starts stagnating in the 500-750s range. Looking at such a loss trajectory, it may be tempting to conclude
running the attack till the local loss converges, should be a good heuristic for knowing when the target ASR will be highest. However, inspecting the case of an adversarially-robust target model (\autoref{fig:local_metrics_track}-b) disabuses us of this notion---ASR peaks at around ten iterations, while the local loss keeps increasing and converging until the very end of attack execution. It is not surprising that local loss continues to converge, since this is what the attacks optimize for while computing gradients, and this may not necessarily align well enough with the target model.

The fact that ASR for the target model keeps increasing significantly even after the attack succeeds for local models is intriguing and a challenge unique to black-box attacks. While this goes hand in hand with the suggestion to evaluate attacks for longer iterations, it raises the question of knowing when the attack running locally should be stopped to maximize ASR for the target model.

\begin{mdframed}
\textbf{Recommendation}: Do not rely on attack success or loss on local models as a metric to stop optimization. Developing metrics that can help predict optimal target ASR is a direction for future work.
\end{mdframed}

\subsubsection{Harder settings} \label{sec:harder_settings}
Since almost all attacks against standard models with sufficient perturbation budget achieve nearly 100\% attack success, there is limited room for improvement. However, attacks in harder settings (\autoref{fig:eps16_vs_eps8_vs_adv}) can be much less effective (\eg $<60\%$ ASR when perturbation budget is halved to 8/255) and can demonstrate different trends in relative performance. For example, against an adversarially trained target model, the least and most performant attacks differ by as much as ${\sim}30\%$ in their ASR (similar trends hold for the targeted setting). Although attacks like \smimifgsm\ seem to perform well across all settings, this is indeed a posterior observation that can only be verified for a new attack when it is evaluated across diverse and hard settings and, in fact, does not hold for other hard settings like IncRes-v2$_\text{ens}$ target models (\smimifgsm\ is out-performed by \vmifgsm\ and \vnifgsm, \autoref{fig:appendix_vgg_resnet_ens} in the Appendix).

\begin{mdframed}
\textbf{Recommendation}: When evaluating and comparing attacks, researchers should include harder attack settings, such as targeted attacks, low perturbation budgets, and adversarially robust target models.
\end{mdframed}

\vspace{1ex}

\section{Discussion}
\label{sec:discussion}

We highlight our key findings, discuss their implications, and make recommendations for future research. We also identify the limitations of this work.

\shortsection{Many Interesting Settings Underexplored}
Categorizing attacks from the literature uncovers how several threat models have close to little or no research dedicated to those specific settings (\autoref{sec:unexplored areas}) despite these areas being some of the most relevant to practical attacks---most model APIs return \topk\ scores (not full confidence vector) and the availability of abundant data from non-overlapping distributions is possible via the Internet, yet both of these settings have hardly any research. We also identify the utility of orthogonal yet useful fields in ML security, such as model extraction and model inversion, and how they can be utilized under certain threat models to boost the performance of black-box attacks (\autoref{sec:new direction}). Future research should focus on developing specific attacks for these unexplored but important and interesting settings.

\shortsection{Careful Evaluation Matters}
Even within well-explored threat spaces, researchers often compare proposed attacks with baselines that require different (and often \revision{more restrictive}) assumptions over the adversary's capabilities, and in settings that are easy enough that all attacks work well. We show how small tweaks to adapt existing methods to utilize the available knowledge fully can strengthen the baselines and outperform the proposed attacks (\autoref{sec:stronger_baselines}).
Additionally, several attacks focus on the untargeted setting where most attacks already achieve near-perfect ASR, instead of harder settings such as targeted attacks and adversarially robust target models, where attack performance trends can change drastically.
We implore researchers to conduct evaluations in settings where differences matter, and to either use state-of-the-art baselines from the same threat space or to adapt baselines to utilize assumed knowledge.

\shortsection{Evaluate Attacks under Well-motivated Constraints}
When constraints are imposed on attacks, they should be motivated by realistic adversarial constraints and focus on attack cost.
Our experiments demonstrate how several proposed attacks can benefit from more iterations, yet predicting the optimal number of local attack iterations is non-trivial.
We thus advocate for a shift in paradigm when reporting attack results for adversarial attacks: using time as the equalizing metric for comparing attacks instead of iterations to infer the attack effectiveness better. We also hope our results motivate future work to use a better selection method for choosing the best candidate examples across iterations.

\shortsection{Limitations}
Our analysis of evaluated attacks is focused on image classifiers, which is not a security-critical application. While there are claims that attacks from image classifiers can be adapted to other domains like malware classifiers, there is little evidence that the decisions about which attack to adapt would be based on extensive evaluations in image space.

\section*{Acknowledgements}
This work was partially funded by awards from the National Science Foundation (NSF) SaTC program (Center for Trustworthy Machine Learning, \#1804603), the 
AI Institute for Agent-based Cyber Threat Intelligence and Operation (ACTION) (\#2229876), NSF \#2323105, and NSF \#2325369. 

\bibliographystyle{IEEEtran} %
\bibliography{main}

\clearpage
\onecolumn{}

\appendix
\subsection{Experiments} \label{app:experiments}

\subsubsection{Implementation Details} \label{app:implementation}
Below, we provide details around our experimental setup and evaluations.

\shortsection{Models} For the normal setting, we consider DenseNet201~\cite{huang2017densely}, Inception-v3~\cite{szegedy2016rethinking}, Resnet101~\cite{he2016deep}, and VGG19~\cite{simonyan2014very} as the target models. All of these normally-trained models were used from the Torchvision~\cite{marcel2010torchvision} library. Additionally, we also consider two robust target models: Inc-v3$_{\text{adv}}$~\cite{Alexey2018Adversarial} and IncRes-v2$_{\text{ens}}$~\cite{Alexey2018Adversarial}. For all of our attacks, our local (surrogate) models consist of an ensemble of:  DenseNet-121~\cite{huang2017densely}, Inception-v4~\cite{szegedy2016rethinking}, ResNet-50~\cite{he2016deep}, and VGG-16~\cite{simonyan2014very}, which do not overlap with the target models.

\shortsection{Data}
We randomly sampled 100 images from the ImageNet~\cite{deng2009imagenet} validation set. To avoid confusion between the two definitions of untargeted attacks (flipping the model's prediction, or making the prediction mismatch the ground-truth), we picked these 100 examples such that all target models have 100\% classification accuracy on them. 

\shortsection{Attacks}
Unless explicitly specified otherwise, all attacks are generated under $16/255$ $\ell_\infty$ perturbation budget, and hyper-parameters are adopted from the original papers for each of the attacks. The typical setting of step size $\alpha$ is set as $\epsilon / T$, where $T$ is number of iterations, and is set as \niterstargeted for targeted attacks, and 10 for untargeted attacks.

\shortsection{Code/Experiments}
We used a batch-size of 5 across all transfer attack experiments to make sure that all attacks (given their varying GPU memory requirements) can fit on the GPU for any given batch. All of our experiments we performed on a 2 CPU, 8-core (2 threads/core) CPU, with 64GB RAM and an Nvidia GTX1080Ti server with 11GB memory, running on Ubuntu Server 22.04. All of our attacks were implemented using PyTorch 1.12.1, running on Python 3.7.13. We exclusively run one experiment at a time on the machine while, although time consuming, helps calculate accurate runtime estimates of attacks without potential fluctuations or slowdowns because of other jobs possibly running on the same machine.

\subsubsection{Attacks Evaluated in This Paper}
\label{sec:evaluated attack background}
Below, we provide brief details about the attacks used for evaluations in \Cref{sec:new_eval}.

\shortsection{Non-interactive Transfer Attacks} %
Fast Gradient Sign Method (\fgsm) \cite{goodfellow2014explaining} generates input perturbation by adding noise in the direction of the sign of gradient of the loss with respect to the input image. I-FGSM (Iterative FGSM) \cite{kurakin2016adversarial} is an iterative version of FGSM that applies the FGSM with smaller step size for multiple iterations and strengthens the effectiveness. \ifgsm\ also becomes the building block of stronger attacks incorporate additional information. For input augmentation methods, \admixfgsm~\cite{wang2021admix} augments the input of \ifgsm\ by adding a small patch from other images. ODS-FGSM \cite{tashiro2020diversity} introduces a sampling strategy for the generated adversarial examples to prioritize diversity in the target model's outputs and improves transferability. The rest of the described attacks enhance the performance of \ifgsm\ with gradient stabilization. \mifgsm\ (Momentum Iterative FGSM) \cite{dong2018boosting} enhances \ifgsm\ by incorporating momentum in gradient calculation while \nifgsm~\cite{lin2019nesterov} uses Nesterov accelerated gradient for \ifgsm\ to effectively look ahead and improve performance. VMI-FGSM~\cite{wang2021enhancing} and \vnifgsm~\cite{wang2021enhancing} respectively further stabilize the \mifgsm\ and \nifgsm\ method by incorporating variance of previous gradients. \smifgsm \cite{wang2022enhancing} considers the (spatial) context gradient information from different regions of the image for stabilization while \smimifgsm\ from the same paper further augments it by adding temporal momentum. \emifgsm~ \cite{wang2021boosting} considers the average gradient of data points sampled in the gradient direction from previous iterations.

\shortsection{Query-based Interactive Attacks}
Bayesian optimization with perturbation sampling from a low dimensional space is leveraged to improve the query efficiency of black-box attacks in the low-query regime, for both the full-score (complete confidence vector) \cite{ru2019bayesopt} and the hard-label settings \cite{shukla2021simple}. The bayesian optimization based attacks can be efficient in the low-query regime as it judiciously chooses the next sample to query based on a proper modeling of the adversarial space distributed around the victim image. However, this attack cannot scale to larger number of queries because the associated Gaussian process will need to maintain a very large kernel matrix, and make the attack extremely slow to optimize and consume huge memory at high number of queries. Some efficient random search based strategies are also proposed for the full-score \cite{andriushchenko2020square} and hard-label \cite{chen2020rays} attacks. Although these attacks are not particularly designed for the low-query regime, they are very efficient to run locally and also shows competitive attack success rate in different query regimes (especially for very high number of queries).

\subsubsection{\topk\ Adaptation Details}
\label{app:topk adaptation}

For untargeted attacks, full-score attacks can be applied directly to the \topk\ setting---most of these attacks only require the prediction score of the ground-truth class, which is always available as the top-1 prediction score except for inputs for which the attack is successful. The setting of targeted attacks is thus much more interesting since the target class may not be included in the \topk\ scores. As an illustration of adapting an attack to this setting, we adapt the Square Attack to the \topk\ targeted attack setting. We call this attack \topksquare.

The \topk\ version of the NES attack (NES: \topk) modifies the original version that operates with complete prediction vector by starting from a random image of the target class (instead of the original seed in the original version), and leverages estimated gradients to gradually reduce the perturbation distance with respect to the original image while still maintaining the class prediction. This way, the confidence score of the target class is guaranteed to be in the \topk\ predictions. We speculate that this idea can also be used to adapt the state-of-the-art Square Attack\cite{andriushchenko2020square}
by starting the attack with a random image of the target class and using corresponding perturbation generation methods to generate perturbed inputs that gradually get closer to the original seed while the target class is still in the \topk\ predictions. However, using the same fixed threshold on the loss function to decide when to start reducing the perturbation size, as done in NES: \topk, does not work for the Square Attack and makes the attack even less ineffective. We solve this by designing a dynamic scheduler that reduces a relatively small threshold (initially $1$ in our experiments) by half if the attack is not successful in finding useful perturbations with reduced size for 10 consecutive iterations, and make the attack successful in generating useful adversarial examples. 
\subsection{Transfer Attacks} \label{app:transfer_attacks}

We provide additional results of transfer attacks on other target models (not covered in the main paper) in \autoref{fig:appendix_vgg_resnet_ens}. The overall findings still support the main claims made in \Cref{sec:new_eval}.

\begin{figure*}[h]
    \includegraphics[width=\textwidth]{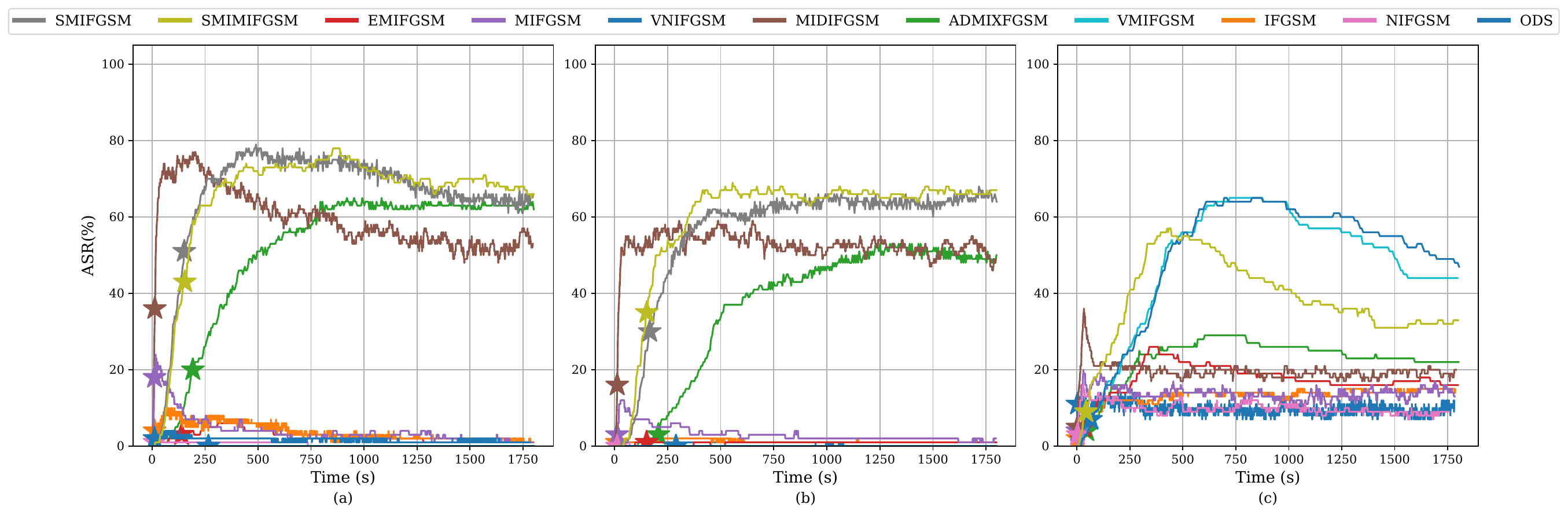}
    \caption{ASR (y-axis) for various attacks varying across time: targeted attacks for VGG19 (a) and Resnet101 (b), and untargeted attacks for IncRes-v2$_{\text{ens}}$ (c). ASR at each iteration is computed using adversarial examples at that iteration. ASR at representative (\niterstargeted for targeted, 10 for untargeted) are marked with $\star$ for each attack. Note that although \smimifgsm\ seems to outperform other attacks in most settings, it is outperformed by \vmifgsm\ and \vnifgsm\ for the case of IncRes-v2$_{\text{ens}}$ (c). ASR at each iteration is computed using adversarial ex, further supporting our argument for evaluation under hard and diverse settings.}
    \label{fig:appendix_vgg_resnet_ens}
\end{figure*}
\subsection{Query-based Attacks} \label{sec:runtime_query_exps}
Query-based attacks compare attacks by tracking ASR as queries are progressively submitted to the target model. Query cost is an important factor, as each query may incur a financial cost \cite{guo2019simple} as well as a risk of detection \cite{chen2020stateful}. However, for resource constrained adversaries or situations where API costs are not a major issue (e.g., model hosted in secure enclave), the local computational cost (runtime) may be a higher priority for attackers. Adversaries in such scenarios might prefer attacks that are efficient to run locally and also require fewer queries. 

\begin{figure}
    \includegraphics[width=\textwidth]{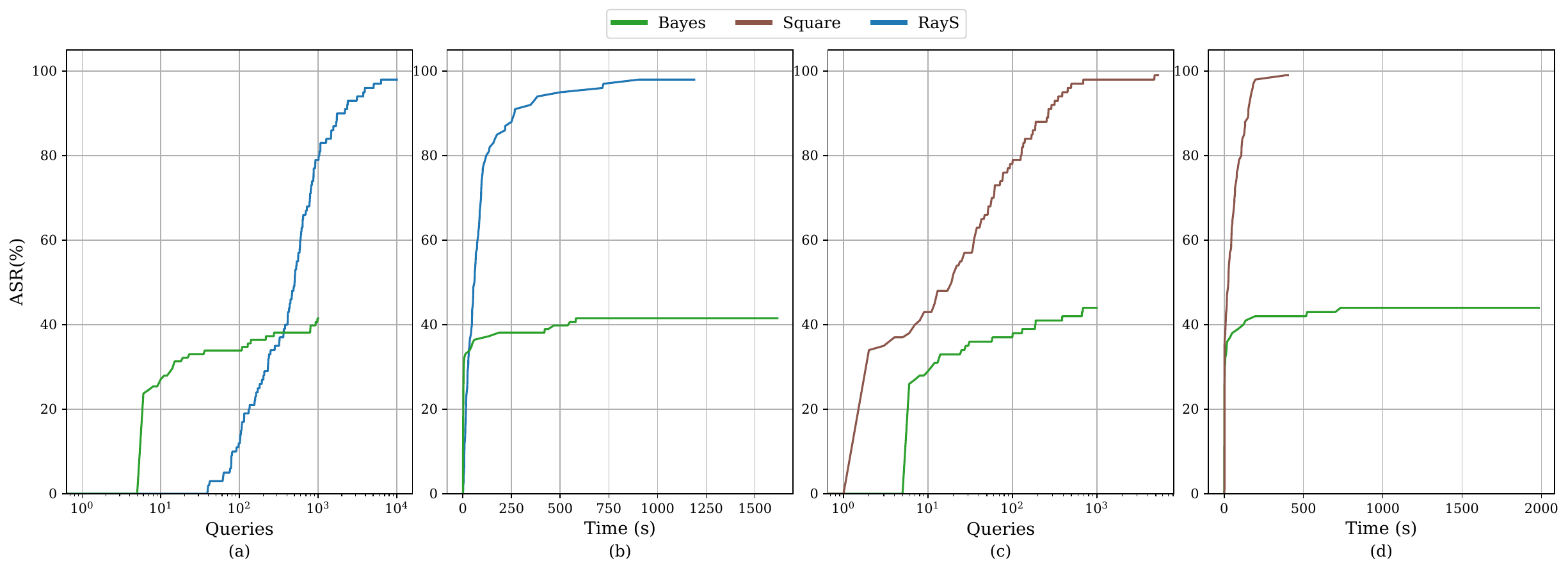}
    \caption{ASR (y-axis) for various query-based untargeted attacks under the hard-label (a, b) and full-score setting (c, d), for Inceptionv3 target model, varying across queries (a, c) and time (b, d). ASR at each iteration is computed using adversarial examples at that number of queries.}
    \label{fig:query_attacks}
\end{figure}

In this section, we compare the bayesian optimization based attack for the hard-label setting (BayesOpt) \cite{shukla2021simple} to the locally efficient RayS attack \cite{chen2020rays}. We choose these two attacks because the first attack achieves state-of-the-art performance in the low-query regime at the cost of high local runtime, while the latter achieves best performance in larger queries and is highly efficient locally. We will use these two attacks to demonstrate how the effectiveness of the attacks can change when the focus of the adversary shifts from the query cost to the local runtime cost. A secondary purpose is to check if BayesOpt attack is still the best in the low-query regime, as the BayesOpt is not compared to RayS in the original paper, despite RayS being published a year before BayesOpt at the same conference. We run untargeted attacks against Inception V3 model and set the query limit to 1,000 for the BayesOpt attack \cite{shukla2021simple} and 10,000 for the RayS attack \cite{chen2020rays}, all consistent with the respective original papers (we do not include targeted attacks since we could not get the BayesOpt attack to successfully generate adversarial examples in the targeted setting within the query limit). 

\autoref{fig:query_attacks}-a shows that the BayesOpt attack still achieves better results in the low query regime, by showing that the ASR is consistently higher than the RayS baseline. This confirms that the BayesOpt attack still achieves better performance in terms of attack success for low numbers of queries. However, when we solely measure the local runtime as the metric (\autoref{fig:query_attacks}-b), the attacks proposed for the efficient attacks with sufficient queries achieve significantly higher attack success rate. Therefore, an attacker with more focus on the local cost might opt for RayS over BayesOpt in practice. Some might argue that the runtime of both attacks on a significant fraction of images are close to 0s. This is because these fraction of seeds are indeed very easy to attack and simple addition of random noise (or adding noises for a few queries) can lead to successful untargeted attacks. This can also be validated by the ASR ($\approx$ 35\%) of Square Attack \cite{andriushchenko2020square} at 1 query, which adds a random noise without receiving the feedback from the target model.  
\revision{Different adversaries under different settings can have different priorities, such as avoiding discovery (minimizing number of queries), or wanting to be computationally efficient (minimizing local runtime). This difference in priorities, along with the demonstrated difference in attack trends, is exactly why it is important to include both kinds of metrics, instead of solely relying on the query based metric, in the future evaluation of query-based attacks.} 

We also repeated the same experiment in the setting of complete confidence vector, where we used the complete confidence score version of the BayesOpt attack using their corresponding implementation \cite{shukla2021simple}, and compared to the state-of-the-art locally efficient Square Attack \cite{andriushchenko2020square}. We note that, there also exists another bayesian optimiation attack \cite{ru2019bayesopt} that is reported to have even higher attack success than the results we obtained by running the BayesOpt attack above. However, the provided code by Ru et al.~\cite{ru2019bayesopt} runs extremely slowly (due to large number of CPU computations) and the attack was also not successful. The authors were also not responsive to our inquiries on possible ways to replicate their results. Therefore, we opt to use the results from the BayesOpt attack mentioned. The results are given in (\autoref{fig:query_attacks}-c,d). We can see that, the locally efficient Square Attack is more efficient than the BayesOpt attack using the both metrics on the number of queries and the local runtime. This shows that, when accessing the complete confidence vector from the target, attacks explicitly proposed for the low-query regime does not seem to be the best option when compared to a more recent baseline, and encourage future research to pick the more competitive baselines for comparison.

\clearpage
\newpage

\end{document}